\documentclass[fleqn]{article}
\usepackage[left=2.5cm,right=2.5cm,top=1.5cm,bottom=1.5cm]{geometry}
\usepackage{amsmath}
\usepackage{float}
\usepackage{graphicx}
\usepackage{amssymb}
\usepackage{multirow,booktabs}
\usepackage[authoryear]{natbib}
\usepackage{authblk} 
\usepackage{float} 
\usepackage{subcaption} 
\usepackage{enumitem}
\usepackage{comment}
\usepackage{newtxtext,newtxmath}
\usepackage{tabularx}
\usepackage{xcolor}
\usepackage[
    colorlinks=true,
    linkcolor=black,
    citecolor=black,
    urlcolor=black,
    filecolor=black,
    pdfborder={0 0 0}
]{hyperref}
\usepackage{pifont}
\usepackage{makecell}
\usepackage[titletoc,title]{appendix}

\newcommand{\Ceqref}[1]{Equation~\eqref{#1}}

\newcounter{remark}
\newenvironment{Remark}{%
  \refstepcounter{remark}%
  \par\noindent\textbf{Remark~\theremark.}\ }{\par}

\newcounter{assumption}
\newenvironment{Assumption}{%
  \refstepcounter{assumption}%
  \par\noindent\textbf{Assumption~\theassumption.}\ }{\par}
  
\title{Analytical modeling of a stop-less modular bus line: Optimization, feasibility, and economies of scale}

\author[1,*]{Haoran Zhao}
\author[2]{Neema Nassir}
\author[1]{Andres Fielbaum}

\affil[1]{\small School of Civil Engineering, The University of Sydney, NSW 2006, Australia}
\affil[2]{\small Department of Infrastructure Engineering, The University of Melbourne, VIC 3010, Australia}

\affil[*]{\small Corresponding author email address: haoran.zhao@sydney.edu.au}

\date{}
\begin{document}
\maketitle

\begin{abstract}
    Conventional bus services often struggle with inefficiencies including prolonged dwell times at heavily used stops, especially for through passengers. A stop-less autonomous modular bus service (SLAM) has been proposed to reduce dwell times by decoupling the front pod to serve stops and then coupling it to the next bus. However, the optimal service design and feasibility region remain underexplored, despite their importance for planning and deployment. We propose an analytical optimization model that characterizes the optimal design, feasibility conditions, and sources of scale economies.
    
    Three novel constraints distinguish SLAM from conventional bus services: (i) a minimum headway to ensure sufficient time for decoupling, alighting, boarding, and coupling operations, (ii) a maximum headway to guarantee all passengers arriving within a headway fit in the standby pod,  and (iii) a minimum bus length constraint, requiring at least two pods per bus to run in a SLAM manner. As ridership grows, the optimal design evolves through several regimes, in which headway constraints alternate between slack and binding states, while capacity constraints shift from one active form to another.
    
    Our analysis indicates that, compared with conventional services, SLAM is most suitable at intermediate demand levels: at low demand, the fixed costs of standby pods and the minimum two-pod configuration outweigh the time-saving benefits, whereas at high demand, non-stopping operation becomes infeasible. We further decompose the sources of scale economies into four components: the Mohring effect, through-capacity economies, boarding-capacity economies, and standby-pod costs, identifying under which conditions each of them is present. The numerical results validate the theoretical analysis.
\end{abstract}
\noindent\textbf{Keywords:} stop-less autonomous modular (SLAM) bus; frequency setting; vehicle capacity; continuous approximation; scale economies

\vspace*{-12pt}
\section{Introduction}
With ongoing urbanization, the global urban population is projected to continue rising \citep{chen_updating_2022}. The concentration of people and activities in cities substantially increases travel demand. With limited road capacity, traffic congestion becomes increasingly severe, especially in metropolitan areas. Thus, public transport plays a crucial role in mitigating congestion, since high-capacity services can move more passengers using far fewer vehicles than private cars.

Buses are a critical component of metropolitan public transport, yet conventional services often suffer from rigid operations, resulting in frequent stops and longer travel times. Decades ago, stop-skipping operations and express service alternatives were introduced to reduce passengers’ in-vehicle travel time \citep{vuchic_skipstop_1976,sun_real_2005,liu_bus_2013,larrain_danger_2020}. By serving only high-demand or interchange stops, such limited-stop services have been implemented worldwide. However, passengers at skipped stops must rely on local routes, leading to longer waiting and total travel times.

In recent years, modular bus services have been proposed to serve passenger demand with flexibility. In particular, \cite{khan_no_2025} have proposed a \textit{Stop-less Autonomous Modular} (SLAM) bus service to eliminate the dwell time of buses. The main idea is that at each stop one pod is decoupled from the modular bus to visit the bus stop, while the remaining pods continue without stopping. This design enables through passengers to travel without intermediate stops, which creates a passenger experience similar to a  fully express service from the origin to the destination on the route.

Case study experiments in \citet{khan_no_2025} demonstrate that the SLAM bus service can significantly reduce passengers’ in-vehicle travel time. Nevertheless, in the absence of systematic optimization, SLAM bus operations may exhibit inefficiencies such as underutilized pods in service or overcrowding. An analytical approach is therefore valuable, as it can clarify how the SLAM bus service responds to demand, assess its economic performance, and compare it with conventional buses. Despite its promising advantages, the optimal design of SLAM systems has received little attention.

To address this gap, we develop an analytical model of SLAM service design. The model provides numerous insights into the optimal design of the system, its feasibility constraints, and sources of scale economies. In comparison to conventional buses, we further discuss the advantages and disadvantages of the SLAM service across various demand levels. Subsequently, the model is solved numerically to validate the theoretical results. To the best of our knowledge, this study is among the first to explore the optimal design of the SLAM bus service. Our study makes three main contributions:
\begin{enumerate}
    \item We derive the optimal service frequency and vehicle capacity of SLAM as functions of demand, and characterize a set of mutually exclusive, collectively exhaustive demand regimes governing the optimal design;
    \item We characterize the feasible demand range for SLAM operation and compare its performance with conventional bus service; and
    \item We decompose the sources of scale economies in SLAM operations, providing insight into the conditions for cost-effective service.
\end{enumerate}

The remainder of the paper is organized as follows. Section 2 reviews autonomous modular buses, as well as continuous approximation and single-line models for bus service design in the literature. Section 3 formulates the analytical model for the SLAM bus service and presents the theoretical analysis. Section 4 presents extensive experiments examining the behavior of the optimal design. Section 5 concludes and suggests directions for future research.

\section{Literature Review}
\subsection{Autonomous modular bus}
Public transport faces persistent challenges, including dwell time at stops, inefficient utilization, unreliable headways and low accessibility at first-mile / last-mile. Conventional fixed-route, fixed-capacity bus systems often struggle to adapt to dynamic passenger flows, especially in off-peak hours or in low-density areas.

The concept of modular autonomous vehicles (or \textit{Autonomous Modular Buses} AMBs) has emerged in recent years as a promising approach to address some of these challenges, since AMBs could introduce flexibility, improve service level, and better utilize the capacity \citep{h_lin_big_2025, khan_no_2025, liu_alleviating_2024}. In an AMB system, vehicles are composed of pods that can couple or decouple dynamically during operation to adjust capacity. Such adjustments could respond to demand fluctuation in real time by increasing or decreasing bus capacity, which efficiently utilizes the bus capacity and avoid empty seats and overcrowding. 

In addition to flexible capacity, in-motion transfer is a major benefit of AMBs. Passengers can perform in-motion transfer by moving to the right pod on the bus that goes to their intended destination, rather than alighting and re-boarding to make a transfer connection. In conventional bus systems, passengers whose origin and destination lie on different bus lines must alight at the transfer stop and board buses operating on another bus line. Even worse, this could occur two or three times per trip.  However, several papers have proposed modular bus designs to avoid such transfers by allowing passengers to move from the current pod to the target pods in the modular bus \citep{wu_modular_2021, khan_seamless_2024, cheng_autonomous_2024}. This intra-vehicle transfer eliminates additional boarding, alighting, and waiting time at stops.

Although AMBs have been found to be promising in the above aspects, challenges remain in this system, including how to schedule, route, and couple/decouple modular pods. To address these challenges, the literature proposes a variety of strategies and optimization models. 
For last-mile operations, \citet{tian_operation_2026} developed a two-stage reinforcement-learning and optimization framework to dynamically determine modular vehicle formations and dispatching decisions under spatially heterogeneous demand and limited fleet resources.
\citet{wang_optimizing_2025} proposed an AMB-based hub-and-spoke public transport system for first-mile/last-mile travel, and found that AMBs can reduce travel time and allow passengers to benefit from in-motion transfers. For mitigating bus bunching, \citet{khan_bus_2023} and \citet{liu_alleviating_2024} studied AMB-based control strategies, using splitting to avoid excessively long headways and holding to prevent overly short headways. \citet{khan_bus_2023} used macroscopic simulation to compare a combined splitting-holding strategy with other heuristics, and found that it reduces the average passenger travel cost and keeps the bunching overhead below 10\%. For modular bus scheduling and formation, \citet{liu_alleviating_2024} developed a continuous approximation model of bus dynamics and treated each bus as an agent in a sequential decision problem, learning when to split and/or hold using a deep reinforcement learning controller. \citet{tian_joint_2023} jointly optimized service schedules and modular vehicle formations under time-dependent demand, considering limited module availability and module rebalancing. \citet{tian_realtime_2025} extended this problem to real-time operations using rolling-horizon control and a learning-based optimization approach. The literature therefore strongly supports the potential of autonomous modular buses to address several operational problems faced by conventional bus systems.

\citet{khan_no_2025} introduced a stop-less autonomous modular (SLAM) bus service, in which the capacity flexibility allows for the near elimination of the dwell time for through passengers. This stop-less operation substantially reduces the travel time. \citet{torres_design_2025} developed a design model for SLAM bus service, optimizing headways and inter-stop distance while assuming a fixed capacity of modular buses.  \citet{yu_service_2025} proposed a bi-level service network design model for modular vehicles, capturing passengers' turning choices and optimized pod circulation patterns. \citet{ye_integrating_2025} developed and tested an integrated personal rapid transit and low-speed autonomous vehicle service, and identified a plan that maintains efficiency while minimizing disruption. However, the capacity design is an important part of the bus service planning, as it jointly with frequency determines the system-level capacity of the service.

\subsection{The single-line model}
Analytical modeling employs explicit mathematical formulations to describe a system’s behavior and the relationships among its key variables. One of its major applications in transport research is the continuous approximation (CA) approach, which provides a stylized analytical framework for modeling public transport systems. It represents discrete operational variables and ridership using continuous functions. The continuous approximation framework offers advantages such as analytical tractability, interpretability, and limited dependence on empirical data, which support generalizable insights, efficient optimization, and sensitivity analysis across a wide range of demand and cost conditions. However, compared to simulation-based approaches, it relies on approximations and yields general insights rather than precise numbers for practical implementation. 

The continuous approximation framework is widely applied in the planning and design of public transport systems \citep{mahmoudi_critical_2024}. One of its applications is the \textit{single-line model}, which represents a single corridor isolated from the broader transit network. This abstraction allows the analysis to focus on the essential components of bus lines, making the model tractable and transparent. Specifically, the single-line model is often used to optimize bus service design. Pioneering this type of analysis, \citet{mohring_optimization_1972} minimizes total system costs and derives the optimal service frequency, which is found to be proportional to the square root of ridership. \citet{jansson_simple_1980} extends this framework to jointly optimize frequency and bus capacity, finding a modified square root rule, and that bus capacity is determined by the capacity constraint to provide sufficient service. Building on decades of development, \citet{jara-diaz_optimal_2017} incorporate demand fluctuations between peak and off-peak hours, minimizing total costs across both periods and obtaining the corresponding optimal frequencies and capacities. \citet{zhang_optimization_2020} formulate a single-line model for on-demand buses, optimizing stop spacing to minimize total passenger travel time, and show that the optimal spacing is highly sensitive to passenger density and to the relative importance of walking versus in-vehicle time. \citet{mei_planning_2021} design a stop-skipping service for a single corridor with continuous demand. By minimizing the total costs - comprising users', operators' and stop-skipping trade-offs, they optimize the stop spacing and the number of stops in each skip-stop bay along the corridor. \citet{ramos_fare_2023} develop a single-line model in which they jointly optimize fare, frequency and inspection rate, and show that high evasion levels justify greater inspection effort and subsidies, while ignoring evasion biases standard single-line design benchmarks. 

The continuous approximation approach has also been applied in studies on modular bus services. \citet{liu_alleviating_2024} use modular buses to alleviate bus bunching, employing continuous approximation to capture the system dynamics arising from the coupling and decoupling of modular units. \citet{chen_continuous_2022} design a modular bus service on a single corridor and use continuous approximation to minimize total costs, including waiting costs, in-vehicle costs, and coupling delays. They obtain the optimal station spacing and service frequency. However, they treat bus capacity as fixed and assume that the number of pods is always sufficient for dispatching.

Despite these developments, little attention has been paid to applying analytical modeling to SLAM bus systems, particularly regarding fleet size and vehicle capacity. This study aims to establish an analytical model for SLAM bus services to determine optimal frequency and bus capacity.

\section{Model formulation}\label{Sec:Model}
\subsection{A recap on the conventional single-line model}\label{sec:ConventionalSingleLine}

Our first aim is to propose an analytical model that enables a theoretical understanding of the optimal SLAM service design and its limits. We note that the SLAM bus service operates on a single line. Therefore, it is useful to first review the conventional \textit{single-line model}.
This is a widely used model, that minimizes the sum of operators' and users' costs. Due to its analytical tractability, it has been used to study optimal bus frequencies, spacing, fares, scale economies, among many other research questions \citep{mohring_optimization_1972, jansson_simple_1980, jara-diaz_optimal_2017,zhang_optimization_2020,c_tang_vehicle_2019,ramos_fare_2023, fielbaum_relationship_2024}. By including a brief summary of this conventional model, the contrasts with SLAM operation will become clearer.

In this conventional model, we consider a line of length $L$ and denote by $X$ the hourly demand. Following most single-line studies, we treat this demand as exogenous and deterministic without stochasticity. We first write the costs as a function of frequency $f$ and bus capacity $K$. According to \citet{jansson_simple_1980,tirachini_economics_2020}, operators' costs $C_O$ can be defined as the fleet size 
$F$ multiplied by the cost of acquiring and operating each bus, $\gamma_0 + \gamma_1 K$. That is, $C_O=F(\gamma_0 + \gamma_1 K)$ where $\gamma_0$ and $\gamma_1$ represent the capacity-independent and capacity-dependent cost components, respectively. The fleet size $F$ is given by the frequency $f$ multiplied by the cycle time $t_c$. The cycle time consists of the time in motion $T$, the time spent boarding and alighting at stops $t X/f$, and time lost due to braking and accelerating at all stops $S \tau$, in which $t$ denotes the average boarding-alighting time per passenger, $\tau$ represents the average time lost due to braking and accelerating per stop, $S$ denotes the number of bus stops. Thus, the cycle time is given by
\begin{equation}
    t_c = T + t \frac{X}{f} + S \tau
\end{equation} and hence the fleet size is determined by 
\begin{equation}
    F=f(T + t \frac{X}{f} + S \tau)
\end{equation}
 Assuming uniform passenger arrivals to stops and regular bus departure headways, users' total costs are estimated by multiplying the ridership $X$ by the sum of average waiting costs $\pi_w/(2f)$ and in-vehicle costs $\pi_v t_cl/L$, where $\pi_w$ and $\pi_v$ are the values of waiting time and in-vehicle time for passengers, and $l/L$ denotes the ratio between the trip length $l$ and the route length $L$. By plugging all estimates into a closed form optimization problem, the problem can be specified as in the following mathematical program.

\begin{equation}\label{obj:trad_single}
    \min_{f,K} \quad \left(\frac{\pi_w}{2f}+\pi_v \frac{l}{L}(T + t \frac{X}{f}+ S \tau)\right)X + f(T + t \frac{X}{f} + S \tau)(\gamma_0 + \gamma_1 K) 
\end{equation}
\begin{equation}\label{con:trad_demand}
    \mbox{s.t.} \quad K \geq \frac{X \rho_{\max}}{f} 
\end{equation}

Where \Ceqref{obj:trad_single} is the objective function including a term that represents passenger-incurred costs (primarily travel and waiting time), which generally decrease with increasing service frequency, and a term that represents operator service and capital costs, which generally increase with service frequency. \Ceqref{con:trad_demand} guarantees that passengers fit into the vehicles, where $X \rho_{\max}$ denotes the maximum passenger load along the line, which will be further detailed when describing the SLAM model (section \ref{sec:modelslam}). Note that the objective function increases with $K$ and therefore \Ceqref{con:trad_demand} is binding. The optimal frequency is obtained by solving the first order conditions, and is given by
\begin{equation}\label{eq:trad_opt_freq}
     f^* = \sqrt{\frac{\frac{X \pi_w}{2} + \pi_v \frac{l}{L}  t X^2 +  \gamma_1 \rho_{\max} t X^2}{(T + S \tau) \gamma_0}}
\end{equation}
The optimal bus capacity is determined by 
\begin{equation}
    K^* = \frac{X \rho_{\max}}{f^*} \label{eq:trad_opt_cap}
\end{equation}
Taking inspiration from this model, we now formulate the service design problem for SLAM bus service.

\subsection{Stop-less autonomous modular (SLAM) bus service}\label{sec:modelslam}
Before introducing the mathematical model for SLAM, let us describe its operation in more detail, by summarizing the system proposed by  \citet{khan_no_2025}. The service operates modular buses on a single line. A modular bus consists of pods with fixed capacity. Crucially, the pods can detach and reattach while moving, allowing the passengers to transfer between pods in the same way as they can walk inside a conventional bus.

\begin{figure}[H]
    \centering
    \begin{subfigure}[H]{0.33\linewidth}
        \centering
        \begin{picture}(0,0)
        \put(-75,0){\textbf{(a)}} 
        \end{picture}
        \includegraphics[width=\linewidth]{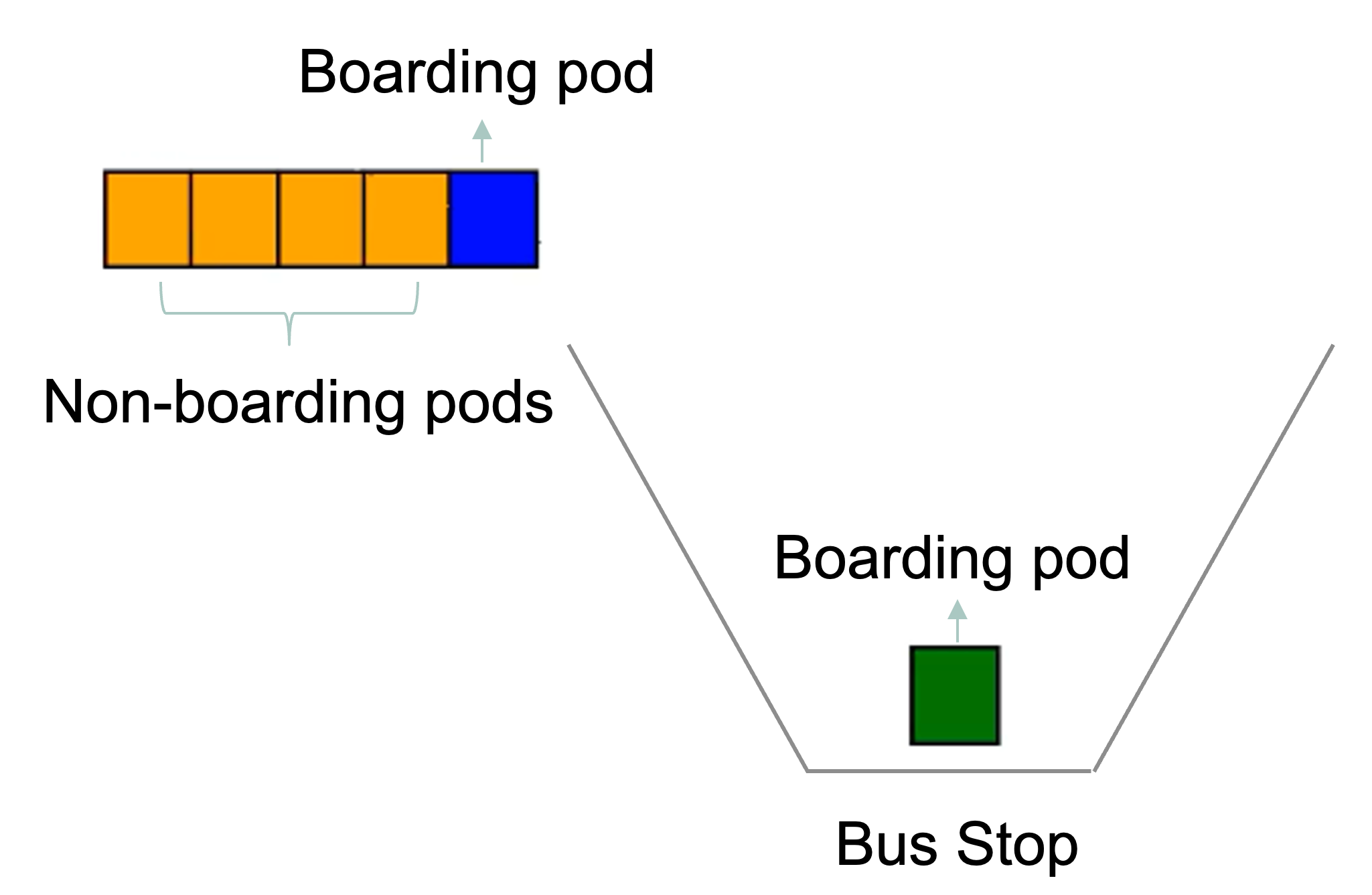}
        \caption{A modular bus, consisting of a boarding pod (blue) and several non-boarding pods (orange), approaches the bus stop}
        \label{fig:SLAM1}
    \end{subfigure}
    \hfill
    \begin{subfigure}[H]{0.3\linewidth}
        \centering
        \begin{picture}(0,0)
        \put(-75,0){\textbf{(b)}} 
        \end{picture}
        \includegraphics[width=\linewidth]{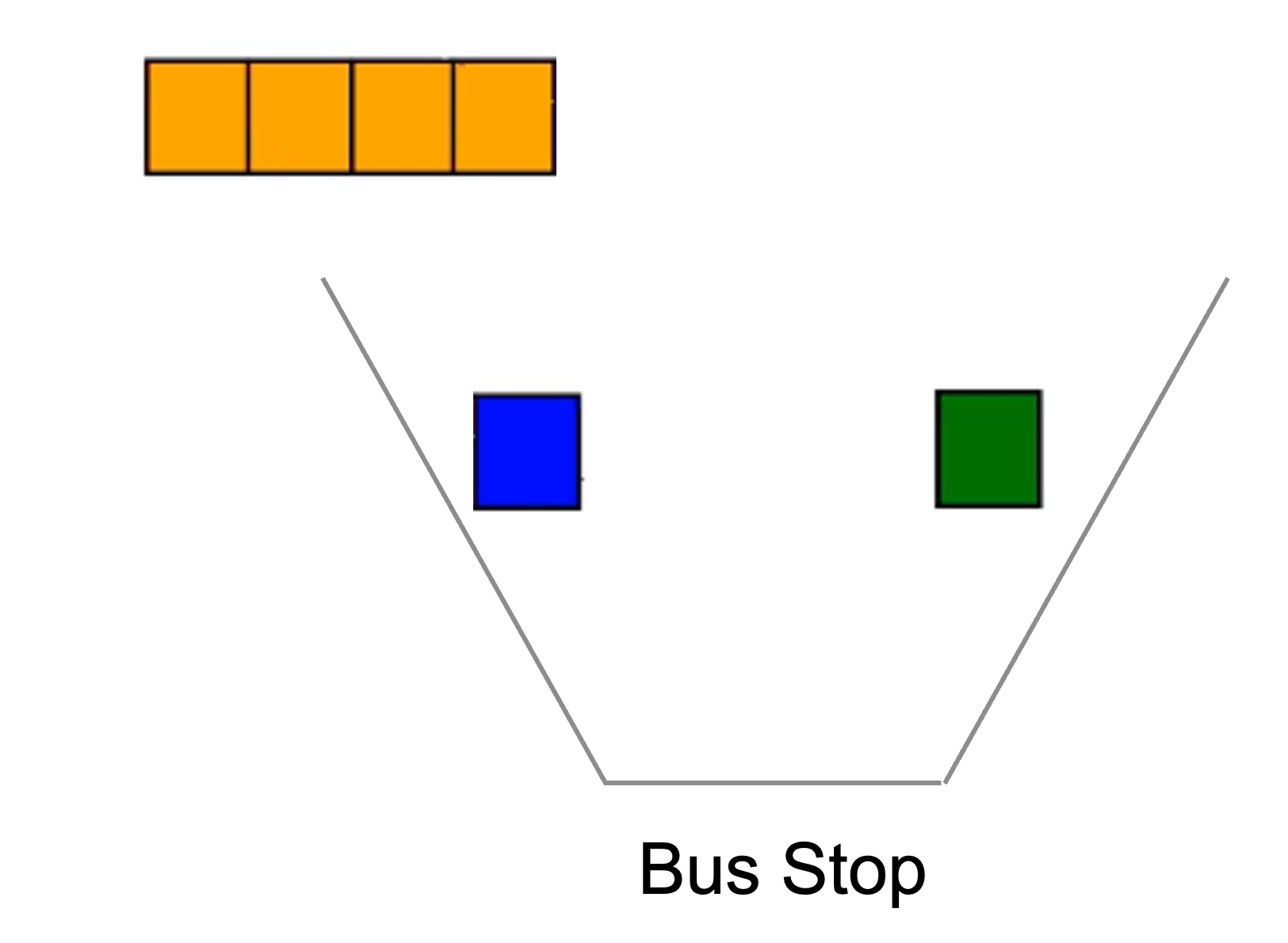}
        \caption{The blue pod detaches from the modular bus carrying alighting passengers, and the green pod leaves the bus stop heading for the modular bus carrying boarding passengers}
        \label{fig:SLAM2}
    \end{subfigure}
    \hfill
    \begin{subfigure}[H]{0.3\linewidth}
        \centering
        \begin{picture}(0,0)
        \put(-75,0){\textbf{(c)}} 
        \end{picture}
        \includegraphics[width=\linewidth]{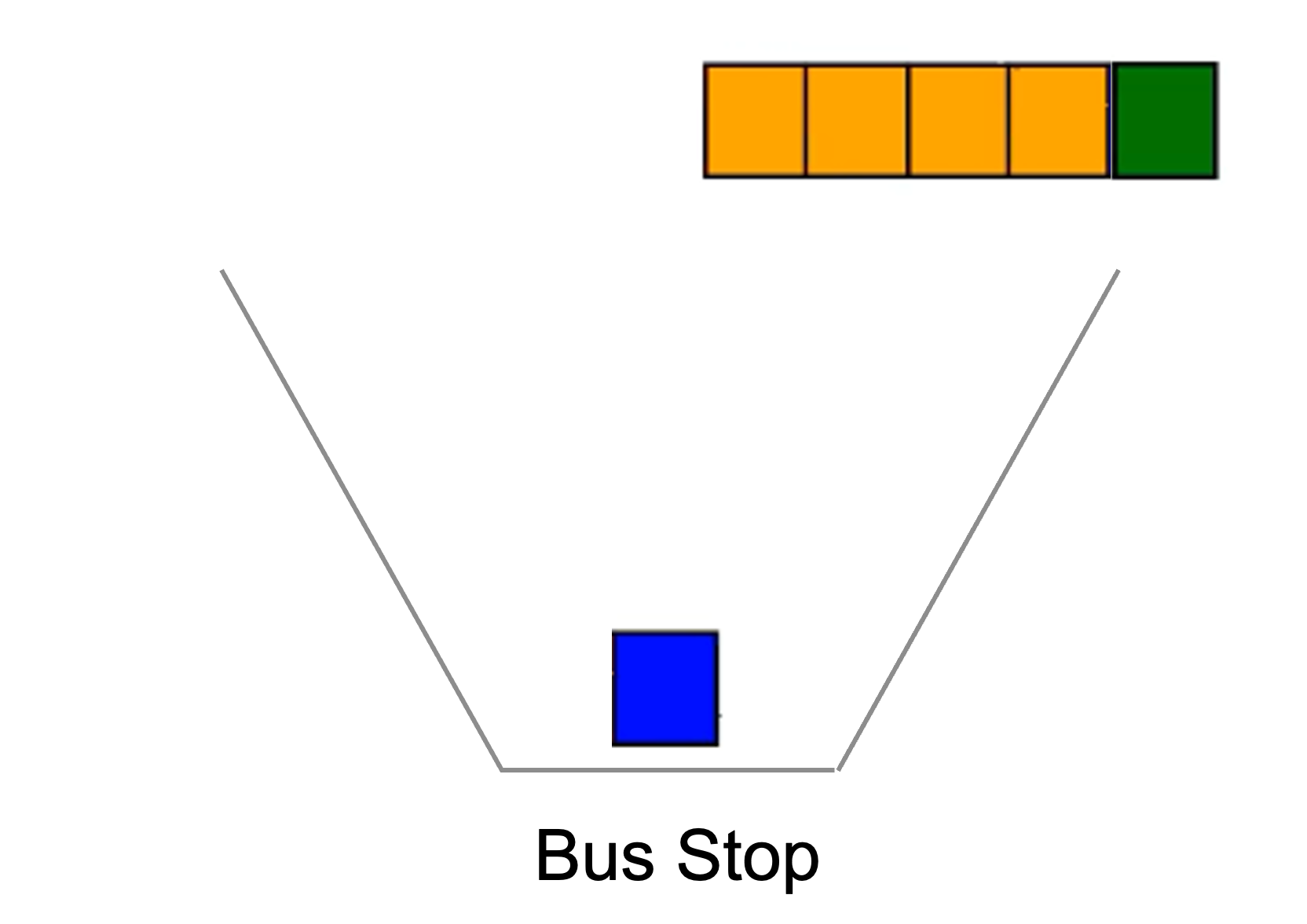}
        \caption{The blue pod arrives at the bus stop, where passengers will alight; after that, new passengers will board the blue pod, and eventually the blue pod will attach to the next arriving bus. The green pod with passengers attaches to the modular bus.}
        \label{fig:SLAM3}
    \end{subfigure}
    \caption{The SLAM bus service moving from left to right performing a non-stopping operation}
    \label{fig:SLAM_illu}
\end{figure}

The main idea of SLAM is that buses never halt at stops. Instead, while the bus continues moving, the front pod detaches for the passengers to alight, and a standby pod at the station, already loaded by boarding passengers, connects to the front of the bus until the next stop. The detached pod will now become the standby pod, and attach to the following bus when it arrives. Passengers intending to alight must move to the front pod, and boarding passengers must enter the standby pod and wait for it to connect to the moving bus. Each front pod of every modular bus and the standby pods at bus stops are defined as \textit{Boarding pods}. \textit{Non-boarding pods} are pods that never halt at stops. Note that only one boarding pod will be decoupled at each stop, which makes the operating pattern easier for passengers to follow.  The system is illustrated in Figure \ref{fig:SLAM_illu}. 
 
\subsection{An analytical model for SLAM}

\subsubsection{Formulation of the model}
Having described the operation of the SLAM bus service, we now formulate an optimization model for its fleet size and bus capacity. Frequency and number of pods per bus are design variables, denoted by $f$ and $P$, respectively. $P$ includes the front pod (note that after the front pod is detached, a bus will have $P-1$ pods till the new pod is reattached). Given that pods have identical capacity of $K_P$ in our model (assumed to be exogenous), the bus capacity increases linearly with the number of pods per bus, as $K = P K_P$. We denote by $\gamma$ the cost per pod. The total demand for the entire line is $X$, and the demand going from stop $i$ to stop $j$ is denoted by $x_{ij}$. The number of stops is $S$. The notation used in the SLAM model is summarized in Table~\ref{tab:notation_slam}.

\begin{table}[H]
    \centering
    \begin{tabularx}{\textwidth}{p{0.18\textwidth} X}
        \hline
        \textbf{Notation} & \textbf{Interpretation} \\
        \hline
        $X$ & Passenger demand on the entire line during one hour. \\
        $f$ & Number of  modular buses dispatched per hour. \\
        $P$ & Total number of modular pods in a bus, including the boarding pod. \\
        $K_P$ & Capacity of a single pod. \\
        $t_c$ & Time required for a bus to complete one cycle. \\
        $S$ & Total number of stops on the line. \\
        $\rho_{\max}$ & Maximum share of hourly demand carried on any segment of the line. \\
        $\phi_{\max}$ & Maximum share of hourly demand boarding or alighting at any single stop. \\
        $\gamma$ & Capital and operating cost of one pod. \\
        $\theta$ & Time to perform the non-stopping operations at a stop. \\
        $t$ & Average time each passenger needs to board or alight. \\
        $\pi_w$ & Passengers' unit value of waiting time. \\
        $\pi_v$ & Passengers' unit value of in-vehicle travel time. \\
        $l$ & Mean passenger trip distance on the line. \\
        $L$ & Total length of the corridor served. \\
        $x_{ij}$ & Hourly passenger demand traveling from stop $i$ to stop $j$. \\
        \hline
    \end{tabularx}
    \caption{Notation used in the SLAM model.}
    \label{tab:notation_slam}
\end{table}

The model is therefore formulated as follows: 

\begin{equation}\label{eq:core_obj}
\min_{f,\,P} \; \left( \frac{\pi_w}{2f} + \pi_v \frac{l}{L} t_c\right) X + (f t_c P + S)\gamma 
\end{equation}
\begin{equation}
\mbox{s.t.} \quad\frac{1}{f} \geq 2 K_P t + \theta\label{con:minimal_hdway_alight}
\end{equation}
\begin{equation}
\phantom{\mbox{s.t.} \quad} \frac{K_P}{X \phi_{\max}} \geq \frac{1}{f}  \label{con:maximal_hdway_pax}
\end{equation}
\begin{equation}\label{con:cap_con}
\phantom{\mbox{s.t.} \quad} K_P \geq \frac{X \rho_{\max}}{f (P-1)}
\end{equation}
\begin{equation}
\phantom{\mbox{s.t.} \quad}  P \geq 2
\label{con:minimum_length}
\end{equation}

We now explain each of the equations and introduce the additional notation. The objective is to minimize the sum of users' and operators' costs as \Ceqref{eq:core_obj}, which resembles exactly the conventional model described in section \ref{sec:ConventionalSingleLine}. The main novelty is that, for SLAM bus service, the cycle time contains no boarding-alighting time due to the non-stopping operation. Therefore, the cycle time equals the time in motion, i.e.,  $t_c = T$. On the other hand, besides the $ft_c$ buses, each of them having $P$ pods, we also have one pod per stop, which is why we have the $(+S)$ term in \Ceqref{eq:core_obj}. Note that we treat $P$ as if it was a continuous variable. We note that as $P$ defines the number of pods per bus, it is by nature discrete. This simplification is included for tractability purposes and further discussed below. 


In the optimization model, the SLAM bus service is subject to the following constraints:


\hspace*{2em}\noindent \textbf{a. Minimum headway constraint} (\Ceqref{con:minimal_hdway_alight}): The front pod detaches from the main bus, goes to the stop where passengers will board and alight, and then move towards the new bus to attach there. The boarding and alighting process can take up to $2K_P t$, as $K_P$ passengers might alight and then $K_P$ passengers can board\footnote{As pods are small, we assume that there is just one door so alighting and boarding occur sequentially.}. The total time for attaching, moving, and reattaching, is denoted by $\theta$. On the other hand, the new bus will arrive after one headway $1/f$. Therefore, this constraint ensures that the boarding pod will have sufficient time to do all of its operations before reattaching to the next bus.

We note that this constraint is specific to SLAM operation. conventional buses will wait at the stop until all passengers board and alight. We call the term $2 K_P t+\theta$ the \textit{minimum feasible headway}, remarking that it is independent from the design variables $f$ and $P$.

\hspace*{2em}\noindent \textbf{b. Maximum headway constraint} (\Ceqref{con:maximal_hdway_pax}): If the headway is excessively long, the accumulation of passengers exceeds the capacity of a single pod, making it unable to accommodate all boarding and alighting passengers.
This means that $X\phi_{\max}/f$ must not exceed the pod capacity, where $X\phi_{\max}$ denotes the maximum number of boarding or alighting passengers across all stops. Specifically, $\phi_{\max}$ is computed as: 
\begin{equation} \label{Eq:phi}
    \phi_{\max} = \frac{1}{X}\max\{\max_i \sum_j x_{ij},\max_j \sum_i x_{ij} \}
\end{equation}

The first term in the maximum denotes the maximum boarding ridership among stops, and the second denotes the maximum alighting ridership. In \Ceqref{Eq:phi}, $\sum_j x_{ij}$ and $\sum_i x_{ij}$ are the passengers boarding and alighting at each stop, respectively.

This constraint is also specific to SLAM operation. In conventional buses, there is no distinction between the passengers that fit in the bus and the passengers that can alight at a single stop. We call the term $K_P/(X\phi_{\max})$ the \textit{maximum feasible headway}, noting that it is independent from the design variables $f$ and $P$.

\hspace*{2em}\noindent \textbf{c. Capacity constraint} (\Ceqref{con:cap_con}): 
Each bus offers a capacity $(P-1) K_P$ when crossing a stop, after the front pod detaches and before the new pod arrives. This capacity must be sufficient to serve the maximum passenger load when crossing a stop $X \rho_{\max}/f$. This load is calculated from the origin-destination matrix $x$. Let $i$ denote the upstream stops, $j$ the current stop under consideration, and $k$ downstream stops. The load is obtained by summing the flows from all upstream origins $i<j$ to all downstream destinations $k>j$, and then taking the maximum across $j$:

\begin{equation}\label{eq:load_rate}
    \rho_{\max} = \frac{1}{X}\max_j \sum_{i<j} \sum_{k>j} x_{ik} 
\end{equation}
In \Ceqref{eq:load_rate}, $\sum_{i<j} \sum_{k>j} x_{ik}$ denotes the number of passengers who board before stop $j$ and alight after stop $j$. Taking the maximum over all stops yields the largest passenger load.

This constraint is equivalent to what \Ceqref{con:trad_demand} represents for conventional buses. The main difference is that this constraint might be more binding. This is because when one pod detaches at every stop, the remaining passengers need to fit in a smaller space ($P-1$ pods instead of $P$).

\hspace*{2em}\noindent \textbf{d. Minimum bus length constraint} (\Ceqref{con:minimum_length}): 
SLAM bus service requires a minimum bus length of two for non-stopping operation, with one boarding pod and one non-boarding pod per bus.


Compared to conventional single-line model in Section~\ref{sec:ConventionalSingleLine}, we introduce three novel constraints into the optimization model for the SLAM bus service. First, minimum and maximum headway constraints are not included in the conventional single-line model \citep{jansson_simple_1980}, since the bus can remain at the stop as long as necessary for passengers to board and alight\footnote{Some models include constraints due to physical considerations (how many buses can be at the stop at the same time), or policy aspects (headways cannot be too long). However, those constraints are from a different nature, as they are completely exogenous, and moreover, less restrictive than the constraints for SLAM.}. In contrast to the uncontrolled headway in conventional single-line model, the headway of SLAM bus service is required to be sufficiently long to accommodate boarding and alighting. The minimum headway constraint ensures that all passengers complete boarding and alighting when the next modular bus arrives. The maximum headway constraint prevents the accumulation of passengers from exceeding the capacity of the boarding pod.

Second, both the conventional and SLAM models include constraints ensuring that passengers fit in the buses. In the case of conventional systems, this is a single constraint considering the maximum load. In the SLAM model, we have two constraints: first, the maximum headway constraint guarantees that the pod capacity exceeds the ridership at stations; and the capacity constraint is for the passengers that remain in the non-boarding pods. In other words, just like the bus is split into two functional parts (boarding and non-boarding), the capacity requirement also splits into two separate constraints — one for each part.

Finally, the minimum bus length constraint is required for non-stopping operations. At each stop, the modular bus detaches its front pod and attaches the standby pod; this maneuver is not feasible if the bus consists of only one pod.

Note that although the problem defined by Eqs. \eqref{eq:core_obj}-\eqref{con:minimum_length} is nonlinear in $f$ and $P$, the continuous model can be solved exactly. For any feasible $f$, the objective function increases with $P$, so the optimal $P$ is always the smallest value allowed by the constraints. The problem can therefore be reduced to one-dimensional convex optimization in $f$ under each possible set of active constraints. Solving these cases yields the global optimum.

\subsubsection{Theoretical analysis}
In this section, we analyze the optimization model as follows. First, we explore the feasible region for a non-stopping operation. Within this region, we next discuss the optimal solutions.

\textbf{Feasible region:} Non-stopping operations of SLAM services fail when the ridership is significantly high. We start by noting that the minimum feasible headway is determined by exogenous parameters, whereas the maximum feasible headway decreases with ridership. The non-stopping operation becomes infeasible when the ridership is large enough so that the maximum feasible headway is lower than the minimum feasible headway. For now we assume the problem is feasible:

\begin{Assumption}\label{asm:always_nonstop}
    The frequency bounds are consistent; i.e., the minimum feasible headway does not exceed the maximum feasible headway:
    \begin{equation}\label{Eq:FeasibleSLAMDemand}
        2 K_P t + \theta \le \frac{K_P }{X \phi_{\max}} 
    \end{equation}
\end{Assumption}  Assumption \ref{asm:always_nonstop} ensures a nonempty feasible interval for frequency. We discuss the specifics of this Assumption, as well as how to adapt our model when the assumption is violated, in Remark \ref{remark:full_stop}. Let us now analyze the optimal design, as the total ridership increases, through the following steps. 

\textbf{Step 1: Monotonicity of the objective function in $P$.}
The objective function increases linearly with the number of pods per bus $P$. To find the minimum objective, the optimal solution is achieved through the smallest feasible $P$. 

\textbf{Step 2: Smallest feasible $P$.}
\Ceqref{con:cap_con} and \Ceqref{con:minimum_length} imply that the optimal number of pods per bus is obtained by
\begin{equation}
    P^*(f)=\max\left\{2, \frac{X\rho_{\max}}{f K_P }+1 \right\} \label{eq:opt_pod_num}
\end{equation}
In other words, at least one of these two conditions is binding, and that condition determines the optimal number of pods per bus.

\textbf{Step 3: Case analysis.} We analyze the optimal design variables in the two cases where either \Ceqref{con:cap_con} or \Ceqref{con:minimum_length} is binding. Recall that the analysis is under Assumption \ref{asm:always_nonstop} and its implications will be discussed in the following Case 2.

\vspace{1em}
\textbf{Case 1:} At low demand levels, the minimum bus length constraint is binding ($P=2 > X\rho_{\max}/(f K_P)+1$). In this case, the SLAM service runs a modular bus of two pods for the non-stopping operation. Substituting $P^* = 2$ into the objective, the optimal frequency is given by
\begin{equation}\label{Eq:MinfreqCase1}
    f^* = \min\left \{\frac{1}{2K_P t + \theta} , \quad \max \left \{\sqrt{\frac{\pi_w X}{ 4 t_c \gamma}},  \quad \frac{X \max\{\rho_{\max}, \phi_{\max}\}}{K_P}  \right \} \right\}
\end{equation}

The term $\sqrt{\frac{\pi_w X}{ 4 t_c \gamma}}$ is obtained through the first order conditions on \Ceqref{eq:core_obj} with $P=2$. This term results in an optimal frequency as long as every other feasibility constraints are fulfilled. The term $\frac{1}{2K_P t + \theta}$ guarantees the minimum headway constraint and the term $\frac{X \max\{\rho_{\max}, \phi_{\max}\}}{K_P}$ guarantees the maximum headway constraint. By comparing the three expressions, the demand region where $f^*=\sqrt{\frac{\pi_w X}{ 4 t_c \gamma}}$ is given by the following equation.


\begin{equation}
\label{eq:simplified_sec_term}
\frac{1}{2K_P t + \theta}\ge\sqrt{\frac{\pi_w X}{ 4 t_c \gamma}} \ge \frac{X \max\{\rho_{\max}, \phi_{\max}\}}{K_P} \Rightarrow X \leq \min \left\{\frac{4 t_c \gamma}{(2K_P t + \theta)^2 \pi_w},\frac{\pi_w K_P^2}{4t_c\gamma \max\{\rho^2_{\max},\phi^2_{\max}\}} \right\}
\end{equation}
This inequality implies that if the demand is sufficiently low, buses with two pods would be the optimal platoon size, and with optimal frequency increasing proportional to the square root of the demand. Under such demand conditions, Equation \eqref{con:cap_con} is not binding. In this regime, only the minimum bus length constraint in Equation~\eqref{con:minimum_length} is binding. We refer to this regime as ``Regime 1: Through Idle Capacity (TIC)'', meaning that the system has idle capacity for through passengers throughout its journey, since the capacity constraint is not binding.

We analyze now the case where the optimal frequency is \(f^* = \frac{X \max\{\rho_{\max},\phi_{\max}\}}{K_P}\), as given in \Ceqref{Eq:MinfreqCase1}. This implies that $X$ exceeds the threshold in \Ceqref{eq:simplified_sec_term}. The optimal frequency is obtained when the minimum bus length constraint in Equation~\eqref{con:minimum_length} is binding together with either the capacity constraint in Equation~\eqref{con:cap_con} or the maximum headway constraint in Equation~\eqref{con:maximal_hdway_pax}. We refer to this regime as ``Regime 2: Partially Loaded Small Buses (PLS)''. Let us assume that $\rho_{\max} \geq \phi_{\max}$, which is almost always the case (more people are on the bus than at the stop), non-boarding pods are full at the most loaded segment and the optimal frequency is determined precisely by that load. In this regime, as demand increases, the optimal frequency increases linearly with demand while maintaining the same level of idle capacity at the front pods.

In the other case, if the minimum in Equation \eqref{Eq:MinfreqCase1} is given by $\frac{1}{2K_P t + \theta}$ (i.e., the Minimum Feasible Headway), which means that the minimum headway constraint becomes binding and determines the optimal solution. Once the optimal frequency is determined by the minimum headway constraint, it will remain constant at its maximum level and no longer increase with ridership. 

In summary, for the case $P^*=2$, we find that the frequency initially increases in proportion to the square root of the demand, and may subsequently increase in direct proportion to demand. Moreover, the upper bound on the frequency imposed by the Equation~\eqref{con:minimal_hdway_alight} (the minimum feasible headway), may be reached under these conditions. 

On the other hand, as the ridership increases, the optimal number of pods per bus will exceed its minimum. It is not necessarily true that having two pods is optimal even before $f$ reaches its upper bound. To analyze this, note that when Equation \eqref{con:cap_con} is active, the objective can be simplified as a function of $P$, which is given by
\begin{equation} \label{Eq:CostFunctionOfP}
    C(P)=\left( \frac{\pi_w(P -1)K_P}{2 X\rho_{\max}} + \pi_v \frac{l}{L} t_c\right) X + \left(\frac{X\rho_{\max}}{K_P(P-1)} t_c P + S\right)\gamma
\end{equation}
This is a convex function of $P$, which is minimized through the first order conditions at 
\begin{equation}\label{eq:opt_P_cap_con}
    P^* =  1 + \sqrt{\frac{2 \rho_{\max}^2 t_c \gamma X }{\pi_w \, K_P^{\,2}}}
\end{equation}
If \Ceqref{eq:opt_P_cap_con} is lower than or equal to two, then $P^*=2$ and our analysis above remains valid. But if \Ceqref{eq:opt_P_cap_con} exceeds two, then it is no longer optimal to use two pods. This situation corresponds to Case 2 below, and the associated demand threshold is determined by
\begin{equation}\label{eq:LowerBoundCase2}
    1 + \sqrt{\frac{2 \rho_{\max}^2 t_c \gamma X }{\pi_w \, K_P^{\,2}}} > 2 \Rightarrow X > \frac{\pi_w K_P^2 }{ 2 \rho^2_{\max} t_c \gamma}
\end{equation}

We remark that the assumption of $P$ being continuous (real number) becomes critical in this analysis. If $P$ were treated as an integer variable, then Equation \eqref{Eq:CostFunctionOfP} would remain valid and convex. Therefore, when the optimal $P^*$ given by Equation \eqref{eq:opt_P_cap_con} is non-integer, the optimal integer solution can be obtained by comparing its two integer neighbors. In particular, if it is greater than or equal to three, then we know that Equation \eqref{con:minimum_length} is no longer binding.




\vspace{1em}
\textbf{Case 2:} As shown in the analysis above, when the demand increases, the capacity constraint becomes binding (i.e., $P = X\rho_{\max}/(f K_P)+1 > 2$). In this case, the SLAM service operates modular buses with more than two pods per bus for a substantial demand.

Substituting $P = X\rho_{\max}/(f K_P)+1$ into the objective yields the optimal frequency as follows:

\begin{equation}\label{eq:opt_headway_case2_core}
     f^* = \min \left \{ \frac{1}{2K_P t + \theta}, \quad \max \left \{\sqrt{\frac{\pi_w X}{ 2 t_c \gamma}}, \frac{X \phi_{\max}}{K_P} \right \}\right\}
\end{equation}
The term $\sqrt{\frac{\pi_w X}{ 2 t_c \gamma}}$ is obtained through the first order conditions and represents the optimal frequency to minimize the total costs. When only the capacity constraint is binding, the system enters ``Regime 3: Partially Loaded Large Buses (PLL)''. Accordingly, non-boarding pods are full, whereas boarding pods have idle capacity. Similar to the TIC regime in Case 1, the optimal frequency grows with the square root of ridership but faster. By comparing three terms in \Ceqref{eq:opt_headway_case2_core}, we obtain the demand region for this case given by
\begin{equation}\label{eq:case2_upper_bound_x}
    \frac{1}{2K_P t + \theta} \ge  \sqrt{\frac{\pi_w X}{ 2 t_c \gamma}} \ge \frac{X \phi_{\max}}{K_P} \Rightarrow  X \leq \min \left\{\frac{2 t_c \gamma}{(2K_P t + \theta)^2 \pi_w}, \frac{\pi_w K_P^2}{2 \phi^2_{\max} t_c \gamma} \right\}
\end{equation}
This equation indicates that neither the minimum nor the maximum headway constraint is binding. Note that the lowest demand in this regime is determined by Equation~\eqref{eq:LowerBoundCase2}. However, this regime only persists until reaching the ridership's upper bound in Equation~\eqref{eq:case2_upper_bound_x}.

We now analyze the case when $X$ exceeds the threshold in \Ceqref{eq:case2_upper_bound_x}, in which the minimum is attained at \(\frac{\pi_w K_P^2}{2 \phi^2_{\max} t_c \gamma}\). Both the capacity constraint in Equation~\eqref{con:cap_con} and the maximum headway constraint in Equation~\eqref{con:maximal_hdway_pax} are binding. The binding capacity constraint implies that the non-boarding pods are full, while the binding maximum headway constraint implies that the boarding pods are full. We therefore refer to this regime as ``Regime 4: Fully Loaded Large Buses (FLL)'', where both boarding and non-boarding pods are full at some segments. The optimal frequency is given by
\(f^* = \frac{X \phi_{\max}}{K_P}\).
Since Equation \eqref{con:cap_con} is binding, the optimal number of pods per bus is determined by
\begin{equation}\label{Eq:PRhoPhi}
    P^* = \frac{\rho_{\max}}{\phi_{\max}} + 1
\end{equation}

Note that the optimal number of pods in this regime is a constant determined by the \textit{peak segment-to-stop concentration ratio} as Equation~\eqref{Eq:PRhoPhi}, where $\phi_{\max}$ denotes demand concentration at the busiest stop (Equation~\ref{Eq:phi}) and $\rho_{\max}$ represents demand concentration along the most-loaded segment (Equation~\ref{eq:load_rate}). A necessary condition for this regime to exist is that $\rho_{\max} \ge \phi_{\max}$, as only in this case Equation \eqref{Eq:PRhoPhi} gives a valid result for $P^* \geq 2$. This condition is natural: as passengers travel at least one stop, the maximum passenger load ratio $\rho_{\max}$ is no less than $\phi_{\max}$, since passengers boarding and alighting at stops are drawn from, or feed into, the cumulative load on the bus. In this regime, the optimal frequency increases linearly with ridership, and the optimal number of pods is a constant which is determined by the peak segment-to-stop concentration ratio. 


Otherwise, if the optimal frequency is determined by the minimum feasible headway, the minimum in \Ceqref{eq:case2_upper_bound_x} will be attained at \(\frac{2 t_c \gamma}{(2K_P t + \theta)^2 \pi_w}\). The optimal frequency reaches its upper bound and the optimal number of pods increases linearly with the ridership. This regime is termed the ``Regime 5: Minimum Feasible Headway (MFH)''.

\vspace{1em}
\textbf{Mutual Exclusivity:} Recall that the feasible region of demand for SLAM operations is 
\begin{equation}
    X \le \frac{K_P}{(2 K_P t + \theta) \phi_{\max}} =  \sqrt{ \frac{2 t_c \gamma}{(2K_P t + \theta)^2 \pi_w} \cdot \frac{\pi_w K_P^2}{2 \phi^2_{\max} t_c \gamma} }
\end{equation}
which is exactly the geometric mean of two thresholds for the FLL and MFH regimes. Therefore, before SLAM operations become infeasible, if the FLL regime emerges, the MFH regime would not appear in the optimal solution and the minimum headway constraint would not be binding. As the ridership increases, the system eventually evolves into one of two cases: 1) the frequency increases linearly with the demand, while the number of pods is fixed based on the spatial distribution of demand (the FLL regime); or 2) the frequency is fixed by the minimum headway constraint, and the number of pods increases linearly with demand (the MFH regime)


\paragraph{Summary of the optimal design and analysis of scale economies}
The SLAM bus service may evolve through five possible regimes as ridership increases.

\vspace{1em}
\textbf{Regime 1: Through Idle Capacity (TIC).} For low ridership $X$, the minimal bus length constraint (\Ceqref{con:minimum_length}) is binding, while the capacity constraint (\Ceqref{con:cap_con}) is not. The optimal frequency increases with the square root of ridership, while the number of pods per bus is fixed at its minimum of 2. The demand region for TIC is given by \Ceqref{eq:simplified_sec_term}, i.e.
\begin{equation}\label{Eq:ICRange}
    X \leq \frac{\pi_w K_P^2}{4t_c\gamma \max\{\rho^2_{\max},\phi^2_{\max}\}}
\end{equation}
and the condition for the existence of this regime is given by 
\begin{equation}
    \frac{\pi_w K_P^2}{4t_c\gamma \max\{\rho^2_{\max},\phi^2_{\max}\}} <\frac{4 t_c \gamma}{(2K_P t + \theta)^2 \pi_w}
\end{equation}
where the ridership threshold of the TIC regime is less than that determined by the minimum feasible headway. The total costs at the optimal solution in this case are \begin{equation}\label{Eq:CostsIC}
    C^*_{TIC} = \underbrace{\left(\sqrt{\pi_w t_c \gamma X}+ \pi_v \frac{l}{L} t_c X\right)}_{\substack{\text{users' costs}}}+ \underbrace{\left( \sqrt{\pi_w t_c X \gamma} + S\gamma \right)}_{\substack{\text{operators' costs}}}
\end{equation}

Scale economies emerge in the TIC regime as $C(X)/X$ is a decreasing function. This is achieved through the terms containing $\sqrt{X}$ and the constant term in Equation \eqref{Eq:CostsIC}, where we identify the \textit{Mohring effect}, the \textit{through-capacity economies}, \textit{boarding-capacity economies} and \textit{standby-pod costs}. First, the waiting-time component of users' costs increases sublinearly with the ridership. This aligns with the well-known Mohring effect \citep{mohring_optimization_1972}, which describes that the average waiting costs decline as the ridership increases and operators provide higher frequency services.

In the TIC regime, only the minimum bus length constraint in Equation~\eqref{con:minimum_length} is binding. Through-capacity economies, boarding-capacity economies and standby-pod costs further reinforce scale economies. The first source arises from through passengers, that is, part of the additional through-passenger demand can be accommodated by the idle capacity in non-boarding pods. The second source arises from boarding passengers, i.e., part of the additional boarding demand can be absorbed by the idle capacity in boarding pods, generating boarding-capacity economies. While these two sources might look similar, there is a fundamental difference between them: when demand increases, the capacity for through passengers can be increased either through the number of pods or frequency, whereas for the boarding pods, this can only be done by increasing frequency. Meanwhile, the fixed costs of having $S$ standby pods are distributed among more passengers.

\vspace{1em}
\textbf{Regime 2: Partially Loaded Small Buses (PLS).} As the ridership continues to increase, both constraints become binding (\Ceqref{con:cap_con} and \Ceqref{con:minimum_length}). The optimal frequency increases linearly with the ridership and the optimal number of pods remains fixed at two. The demand region characterizing PLS is given by
\begin{equation}\label{Eq:FSBRange}
     \frac{\pi_w K_P^2 }{ 2 \rho^2_{\max} t_c \gamma} \geq X >\frac{\pi_w K_P^2}{4t_c\gamma \max\{\rho^2_{\max},\phi^2_{\max}\}}
\end{equation}
The condition for the existence of this regime is that the ridership threshold is less than that determined by minimum feasible headway, as given by 
\begin{equation}
    \frac{\pi_w K_P^2 }{ 2 \rho^2_{\max} t_c \gamma} <\frac{4 t_c \gamma}{(2K_P t + \theta)^2 \pi_w}
\end{equation}

The waiting time and the number of standby pods still demonstrate the scale economies. In this case, the minimal total costs are 
\begin{equation}
    C^*_{PLS} = \underbrace{ \frac{\pi_w K_P}{2  \max\{\rho_{\max},\phi_{\max}\}} + \pi_v \frac{l}{L} t_c X}_{\substack{\text{users' costs}}} + \underbrace{ 2 \frac{X \max\{\rho_{\max},\phi_{\max}\}}{K_P} t_c\gamma + S\gamma}_{\text{operators' costs}}
\end{equation}

The frequency increases linearly with $X$ in the PLS regime, but in the TIC regime it increases with the square root of the ridership. This indicates more significant time savings gained by passengers as demand increases in this regime. In addition, the standby-pod term also continues to contribute to scale economies in both regimes. That is, scale economies result from the Mohring effect and the costs of standby pods. 

\vspace{1em}
\textbf{Regime 3: Partially Loaded Large Buses (PLL).} As the ridership grows even further, it might happen that \Ceqref{con:cap_con} remains binding, but \Ceqref{con:minimum_length} is no longer active. In this regime, both the frequency and capacity of the system increase proportional to the square root of the demand. In this case, the demand region for PLL is given by
\begin{equation}\label{Eq:FLBRange}
    \frac{\pi_w K_P^2}{2 \phi^2_{\max} t_c \gamma} \ge X > \frac{\pi_w K_P^2 }{ 2 \rho^2_{\max} t_c \gamma} 
\end{equation}

The condition for the existence of this regime, considering the minimum feasible headway, is given by 
\begin{equation}
    \frac{\pi_w K_P^2}{2 \phi^2_{\max} t_c \gamma} <\frac{2 t_c \gamma}{(2K_P t + \theta)^2 \pi_w}
\end{equation}

Scale economies become less pronounced in this case. The minimal total costs are
\begin{equation}\label{Eq:FLB_OBJ}
    C^*_{PLL} = \underbrace{\sqrt{\frac{ \pi_w t_c \gamma X}{2}}+ \pi_v \frac{l}{L} t_c X }_{\text{users' costs}}  +  \underbrace{  t_c \gamma\frac{ X \rho_{\max}}{K_P} + \sqrt{\frac{\pi_w t_c \gamma X}{ 2 }} + S\gamma }_{\text{operators' costs}}
\end{equation}
Compared to the TIC and PLS regimes, the average waiting costs decrease more slowly (proportional to $(2x)^{-1/2}$ rather than $X^{-1}$ or $x^{-1/2}$ ). This results from a shift from increasing frequency alone to increasing both frequency and capacity. The optimal frequency increases at a slower rate than in the TIC regime, resulting in weaker waiting-time scale economies. In this regime, only the capacity constraint in Equation~\eqref{con:cap_con} is binding. That is, the front pod has idle capacity but the through pods become full at the most loaded segment. The boarding-capacity economies occur  because the boarding pods are not fully occupied, so additional boarding passengers can be accommodated without proportional capacity expansion. They influence total costs through the capacity constraint, generating an additional term in the optimal fleet size that depends on ridership in a sublinear way. Consequently, the scale economies in this regime stem from three mechanisms: the Mohring effect, boarding-capacity economies and standby-pod costs.

\vspace{1em}
\textbf{Regime 4: Fully Loaded Large Buses (FLL)}. When the ridership continues to increase, it might occur that the optimal frequency increases linearly with ridership, but never faster than in the PLS regime. In this regime, both the capacity constraint in Equation~\eqref{con:cap_con} and the maximum headway constraint in Equation~\eqref{con:maximal_hdway_pax} are binding. Jointly imposing these two constraints yields a constant optimal number of pods, determined by the ratio of $\rho_{\max}$ to $\phi_{\max}$. The condition of existence of this regime is $\rho_{\max} > \phi_{\max}$, resulting from the optimal number of pods in Equation~\eqref{Eq:PRhoPhi} exceeding the minimum bus length. 

The demand region of this regime is given by
\begin{equation}\label{Eq:FLBIIRange}
    \frac{K_P}{(2 K_P t + \theta)\phi_{\max}} \ge X > \frac{\pi_w K_P^2}{2 \phi^2_{\max} t_c \gamma}
\end{equation}
considering the upper bound imposed by the feasibility of the non-stopping operations and the lower bound inherited from the preceding regime.
According to the theoretical analysis, the eventual occurrence of the FLL regime implies that the MFH regime does not emerge. Scale economies are more significant than in the PLL regime but weaker than in the PLS regime. The minimal total costs are
\begin{equation}
    C^*_{FLL}= \underbrace{\frac{\pi_w K_P}{2\phi_{\max}}+ \pi_v\frac{l}{L}t_c X}_{\text{users' costs}}\;+\; \underbrace{ \gamma\frac{t_c(\rho_{\max}+\phi_{\max})}{K_P} X +  S \gamma}_{\text{operators' costs}}
\end{equation}
The total waiting time costs are constant in the FLL regime, which is a significant source of scale economies. Although the optimal number of pods per bus is constant, the optimal frequency increases linearly with ridership, and so does the fleet size. Consequently, the linear fleet expansion no longer yields scale economies. The standby-pod costs still contribute to scale economies. In this regime, the sources of scale economies are only Mohring effect and standby-pod costs.

\vspace{1em}
\textbf{Regime 5: Minimum Feasible Headway (MFH)}. Finally, the frequency reaches its maximum, given by the inverse of the minimum feasible headway. The number of pods per bus (bus capacity) increases linearly with the ridership. The demand region for MFH is expressed as
\begin{equation}\label{Eq:MFHRange}
    \frac{K_P}{(2 K_P t + \theta)\phi_{\max}} \ge X >  \frac{2 t_c \gamma}{(2K_P t + \theta)^2 \pi_w}
\end{equation}
Given a fixed frequency at its maximum and a deterministic cycle time, operators are running the same number of modular buses but increasing the number of pods per bus in the MFH regime. In this case, the minimal total costs are
\begin{equation}
    C^*_{MFH}
= \underbrace{\frac{\pi_w}{2} (2 K_P t +\theta )X + \pi_v\frac{l}{L}t_c X}_{\text{users' costs}}+ \underbrace{\gamma \frac{t_c\rho_{\max}}{K_P}X 
+ \gamma\frac{t_c}{2K_P t + \theta}
+ S \gamma }_{\text{operators' costs}}
\end{equation}
The waiting-time term no longer contributes to scale economies as the frequency is now fixed. However, boarding-capacity economies contribute to scale economies in this regime. Given that the minimum headway constraint in Equation~\eqref{con:minimal_hdway_alight} is binding and the frequency is fixed, having more passengers reduces idle capacity in boarding pods. Consistently across five regimes, the standby-pod costs contribute to the scale economies.

As a synthesis of the preceding analysis, Table~\ref{tab:regimes} summarizes the operational regimes, their corresponding ridership ranges, and sources of scale economies. We remark that the five regimes above are possible regimes rather than guaranteed outcomes. Depending on the parameter values, some regimes may not arise. In particular, FLL and MFH are mutually exclusive. Specifically, we have derived the thresholds that mark when each of these regimes begin. If the threshold for the last regime is lower than any of them, the corresponding regimes would not appear. The emergence of regimes in the sequence is governed by the parameters, among which pod capacity is one of the most important factors and is further explored in the experiments. Subsequently, the possible combination of all regimes above are: 
\begin{enumerate}
    \item MFH,
    \item TIC $\rightarrow$ MFH,
    \item TIC $\rightarrow$ PLS $\rightarrow$ MFH,
    \item TIC $\rightarrow$ PLS $\rightarrow$ PLL $\rightarrow$ MFH , or
    \item TIC $\rightarrow$ PLS $\rightarrow$ PLL $\rightarrow$ FLL .
\end{enumerate}

\begin{table}[H]
\centering
\begin{tabularx}{\textwidth}{|X|X|X|X|X|X|X|X|X|}
\hline
\multirow{2}{*}{Regimes} 
& \multirow{2}{*}{$X_{\min}$} 
& \multirow{2}{*}{$X_{\max}$} 
& \multirow{2}{*}{\makecell[l]{Optimal \\ $f$ behavior}} 
& \multirow{2}{*}{\makecell[l]{Optimal \\ $P$ behavior}} 
& \multicolumn{4}{c|}{Sources of scale economies} \\
\cline{6-9}
& & & & & Mohring effect & Through-capacity economies & Boarding-capacity economies & Standby-pod costs \\
\hline
Through Idle Capacity (TIC) & 0 &Eq.\eqref{Eq:ICRange} &Square root of ridership &Minimum &\ding{51} &\ding{51} & \ding{51}&\ding{51} \\
\hline
Partially Loaded Small Buses (PLS) &Eq.\eqref{Eq:FSBRange} &Eq.\eqref{Eq:FSBRange} & Linear with ridership &Minimum &\ding{51} & & &\ding{51} \\
\hline
Partially Loaded Large Buses (PLL) &Eq.\eqref{Eq:FLBRange} &Eq.\eqref{Eq:FLBRange} &Square root of ridership &Capacity constraint &\ding{51} & & \ding{51}&\ding{51} \\
\hline
Fully Loaded Large Buses (FLL) &Eq.\eqref{Eq:FLBIIRange} &Eq.\eqref{Eq:FLBIIRange} &Linear with ridership &Capacity constraint &\ding{51} & & &\ding{51} \\
\hline
Minimum Feasible Headway (MFH) &Eq.\eqref{Eq:MFHRange} &Eq.\eqref{Eq:MFHRange} &Maximum & Capacity constraint& & &\ding{51} &\ding{51} \\
\hline
\end{tabularx}
\caption{SLAM bus service regimes derived from the behavior of the optimal design variables, along with corresponding ridership ranges and sources of scale economies.}
\label{tab:regimes}
\end{table}

{\subsubsection{Feasibility of non-stopping operations}

Maximum and minimum headway constraints together determine the feasibility of non-stopping operations. These two numbers are both derived from different parameters. What happens if there is no feasible range? The original SLAM model by \cite{khan_no_2025} considers the possibility of ``full stops'', that is, where all the pods visit the stop instead of only the boarding pod. 

\vspace{1em}
\begin{Remark} \label{remark:full_stop}
 If the bus does a full stop, then the corresponding stop should be removed from the computation of $\phi_{\max}$ in Eq.~\eqref{Eq:phi}. By doing so, the value of $\phi_{\max}$ is reduced as the maximum is taken over a smaller set of stops. 
    
    Therefore, when Assumption 1 is dropped, our model can be used to identify in which stops the bus needs to do a full stop, namely, whenever the number of hourly passengers boarding or alighting (or both) exceeds $\frac{K_P}{(2K_Pt+\theta)\phi_{\max}}$. To give a more concrete flavor, let us remark that if we take the parameters we use when running numerical experiments (Table \ref{tab:constants_simu} in section \ref{sec:experiments}), the threshold is around 4000 passengers per hour for pods with 6 seats, which is a large value but that could be reached under very high demand conditions. Moreover, as our model assumes a steady demand, the actual threshold should be lower to accommodate for potential random demand fluctuations.
\end{Remark}

The idea of full stops suggests a somewhat binary operation: either only the front pod stops, or all of them. It could be possible to design a system where this decision depends on the stop or is taken dynamically. However, this would imply that the system is significantly harder for the passengers to understand.

\subsection{Comparison between SLAM and conventional bus service}

\textbf{Behavior of the optimal solution}
In the conventional single-line bus model, the optimal frequency obtained from \Ceqref{eq:trad_opt_freq} increases with ridership without bound, denoted by
\begin{equation}\label{Eq:OptFreqTrad}
    f^* = \sqrt{\frac{\frac{X \pi_w}{2} + \pi_v \frac{l}{L}  t X^2 +  \gamma_1 \rho_{\max} t X^2}{(T + S \tau) \gamma_0}} \quad \mbox{such that, as} \quad X \to +\infty,\; f^*\to +\infty
\end{equation}

The optimal bus capacity from \Ceqref{eq:trad_opt_cap} increases with the ridership. Specifically,
\begin{equation}
    K^* = X\rho_{\max} \sqrt{\frac{(T + S \tau) \gamma_0}{\frac{X \pi_w}{2} + \pi_v \frac{l}{L}  t X^2 + t \gamma_1 \rho_{\max}X^2}} = \rho_{\max} \sqrt{\frac{(T + S \tau) \gamma_0}{\frac{ \pi_w}{2X} + \pi_v \frac{l}{L} t + t \gamma_1 \rho_{\max}}} 
\end{equation}
and hence as $X \to +\infty$,
\begin{equation}
     K^* \to \rho_{\max} \sqrt{\frac{(T + S \tau) \gamma_0}{ \pi_v \frac{l}{L} t + t \gamma_1 \rho_{\max} }}
\end{equation}

In the SLAM model, the frequency initially increases with the square root of demand, as in the conventional model. Before the frequency reaches its maximum, as the demand continues to grow, the pattern changes: we observe up to two cycles in which frequency alternates between increasing in proportion to the square root of demand and increasing directly in proportion to demand until being determined by the peak segment-to-stop concentration ratio. Consequently, the regimes of the optimal design are ``Through Idle Capacity (TIC) - Partially Loaded Small Buses (PLS) - Partially Loaded Large Buses (PLL) - Fully Loaded Large Buses (FLL)''. At low demand, bus capacity is fixed at its minimum of 2, and the frequency is proportional to the square-root of the demand. In this regime, the SLAM bus service accommodates the demand by increasing the fleet size of small modular buses.

At higher demands, the SLAM bus service could serve the ridership by small buses with high frequency if the maximum headway is allowed (the ``Fully Loaded Large Buses'' regime). Otherwise, the SLAM bus service operates at its maximum frequency and increases bus length (the ``Minimum Feasible Headway'' regime). The former regime accommodates the increasing ridership by deploying more buses, while the latter does so by increasing the number of pods per bus.



\paragraph{Total Costs Comparison} We compare the SLAM service with conventional bus services across different levels of demand to assess their relative performance.

Under low-demand conditions, the SLAM bus service may be less economical than conventional bus services. As discussed in \ref{appdx:Comp_low_demand}, operators deploy standby pods to enable non-stopping operations, which reduces in-vehicle time for through passengers but incur additional costs for acquiring and operating these pods. However, the benefit of reducing dwell time is outweighed by costs of acquiring and operating additional pods, thereby diminishing the cost advantage of the SLAM bus service. Moreover, SLAM may increase passengers' waiting times relative to conventional bus services. This effect arises from the minimum bus length constraint: if this constraint were removed, the optimal frequency could double during the Through Idle Capacity (TIC) regime. A lower service frequency therefore leads to longer waiting times. It is important to note that SLAM bus services prioritize reductions in in-vehicle time rather than minimizing waiting time.

As the ridership increases, SLAM outperforms the conventional services as long as non-stopping operations remain feasible. With further increase in demand, if the number of waiting passengers is excessively large, all pods within the modular bus would be required to stop. At this point, however, the system does not immediately switch to a fully conventional service at all stops. Instead, the system can decide whether to perform non-stopping operations at every bus stop. This could lead to a hybrid state where both SLAM and conventional service coexist within a service cycle, rather than enforcing a uniform stopping pattern. When non-stopping operations are infeasible at all stops, the service reverts to the conventional bus operation.

In conclusion, the SLAM bus service is the most suitable for intermediate demand levels. The relative performance is illustrated by Figure~\ref{fig:SLAMDemand}. The figure shows a threshold separating the low- and intermediate-demand levels. Below this threshold, conventional buses are more cost-effective than SLAM.  Although this threshold cannot be expressed in a simple closed form, its existence is established in \ref{appdx:Comp_low_demand}.

The threshold between the intermediate- and high-demand levels is determined by 
\begin{equation}
X_{\text{SLAM}}^{\text{max}} = \frac{K_P}{(2 K_P t + \theta) \phi_{\max}}
\end{equation} 
which is derived from Equation~\eqref{Eq:FeasibleSLAMDemand} and represents the maximum demand under which SLAM remains feasible for all stops. Beyond $X_{\text{SLAM}}^{\max}$, modular buses must begin serving some stops with full stopping operations, eventually serving all stops in this manner. 
Accordingly, the low- and intermediate-demand ranges in Figure~\ref{fig:SLAMDemand} cover the five SLAM operating regimes identified in our analysis. The demand threshold $X_{\text{SLAM}}^{\min}$ may arise in one of the identified SLAM operating regimes, depending on the parameter values.

\begin{figure}[H]
    \centering
    \includegraphics[width=\linewidth]{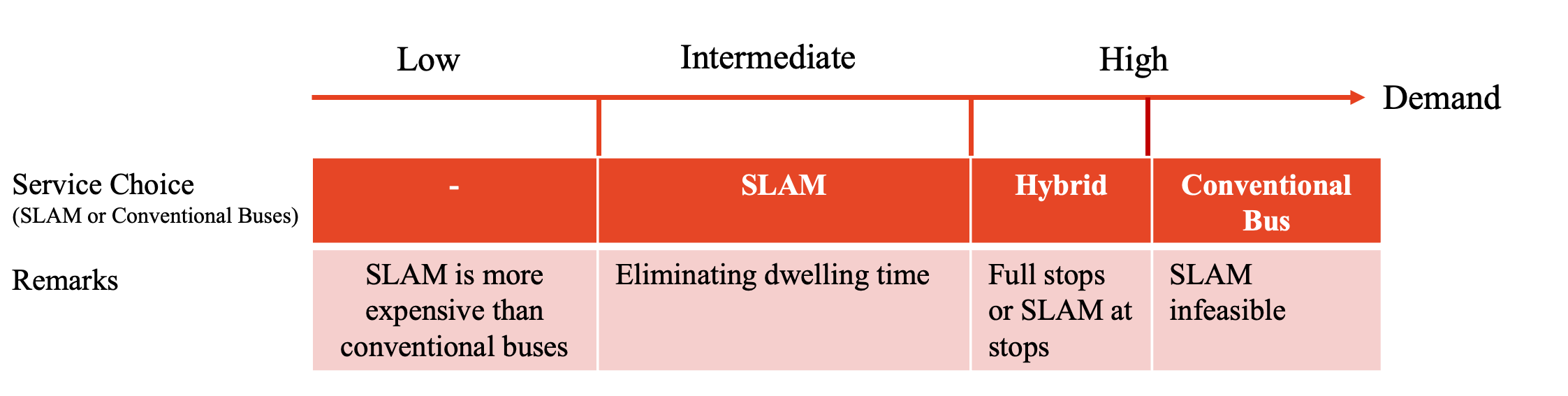}
    \caption{The comparison between SLAM and conventional services under various levels of demand}
    \label{fig:SLAMDemand}
\end{figure}


\section{Experiments}\label{sec:experiments}
In this section, we run numerical experiments to observe the behaviour of the system under realistic values of the parameters. 

\subsection{Settings}
The values for the parameters related to users' costs follow \citet{fielbaum_transit_2018}. The operators' cost values are taken from \citet{tirachini_economics_2020}, as the SLAM bus service is electrified, based on automated driving and saves the cost of driving completely. The values of the operator-cost constants for conventional buses follow \citet{fielbaum_transit_2018}. These values are listed in Table~\ref {tab:constants_simu}.


\begin{table}[htbp]
\begin{tabular*}{\hsize}{@{\extracolsep{\fill}}lllp{10cm}@{}}
\hline
Category & Notation & Value & Description \\
\hline
\multirow{5}{*}{Bus line} 
    & $S$ & 20 (stops) & Number of bus stops for the entire cycle \\
    & $K_P$ & {6 (seats), 16 (seats)} & Pods capacity \\
    & $D$ & 400 (m) & Average distance between bus stops \\
    & $V$ & 20 (km/h) & Average travel speed of pods \\
    & $t$ & 2 (s) & Average boarding and alighting time per passenger \\
    & $\tau$ 
    & 30 (s) 
    & Average time lost due to braking and acceleration per stop \\
    & $\theta$ & 30 (s) & The fixed time cost for coupling and decoupling pods at each stop \\
\hline
\multirow{3}{*}{Demand}
    & $X$ & 1000 (pax/h) & Hourly demand of the bus line \\
    & $\rho_{\max}$ & 0.4 & the ratio of the maximum load to the hourly demand \\
    & $\phi_{\max}$ & 0.1 & \makecell[l]{the ratio of the maximum boarding and alighting \\ passengers to the hourly demand} \\
    & $l$ & 2 (km) & Average travel distance \\
\hline
\multirow{5}{*}{Costs}
    & $\pi_w$ & 4.44 (\$/h) & Time value of waiting time \\
    & $\pi_v$ & 1.48 (\$/h) & Time value of in-vehicle time \\
    & $\gamma$ & \makecell[l]{5.34 (\$)(6 seats pod),\\ 8.04 (\$)(16 seats pod)} & Operators' costs per pod of SLAM\\
    & $\gamma_0$ 
    & 4.02 (\$) 
    & Capacity-independent component of operators' costs for conventional buses \\
    
    & $\gamma_1$ 
    & 0.89 (\$/seat) 
    & Capacity-dependent component of operators' costs for conventional buses  \\
\hline
\end{tabular*}
\caption{Values of constants used in the optimization of SLAM design and comparison with conventional buses.}
\label{tab:constants_simu}
\end{table}


\subsection{SLAM bus service design}
In this subsection, we optimize frequency and the number of pods per bus for the SLAM bus service. In the theoretical analysis, we find five possible regimes of the optimal design: ``Through Idle Capacity'', ''Partially Loaded Small Buses'',``Partially Loaded Large Buses'', ``Fully Loaded Large Buses'', and ``Minimum Feasible Headway''. Depending on the values of constants, three intermediate regimes may not appear, and only one of ``Fully Loaded Large Buses'', and ``Minimum Feasible Headway'' could emerge before the SLAM operation becomes infeasible. To thoroughly illustrate the behavior of the optimal design, we therefore present the optimal design when the pod capacity is 6 seats and 16 seats.

\subsubsection{Case of small pods with capacity of 6 seats}\label{sec:6_seats_case}
We first set the pod capacity to 6 seats. As illustrated in Figure~\ref{fig:opt_small}, the optimal design exhibits four regimes before the SLAM becomes infeasible.

First, the Through Idle Capacity regime persists until the ridership reaches 120 passengers per hour, during which the optimal frequency increases with the square root of the ridership and the number of pods remains at its minimum. Subsequently, the Partially Loaded Small Buses regime emerges, where the optimal frequency increases linearly with the ridership and the number of pods is determined simultaneously by minimum bus length and capacity constraint. 

Next, the Partially Loaded Large Buses regime dominates the feasible region for ridership levels between 240 and 3800 passengers. In this regime, both the optimal frequency and number of pods grow with the square root of the ridership. Finally, before the SLAM operation becomes infeasible, the Fully Loaded Large Buses regime fixes the optimal number of pods at a constant level determined by the peak segment-to-stop concentration ratio ($\rho_{\max} / \phi_{\max}$), while the optimal frequency increases linearly with ridership. 

\begin{figure}[H]
    \centering
    \begin{subfigure}[t]{0.45\linewidth}
        \centering
        \begin{picture}(0,0)
        \end{picture}
        \includegraphics[width=\linewidth]{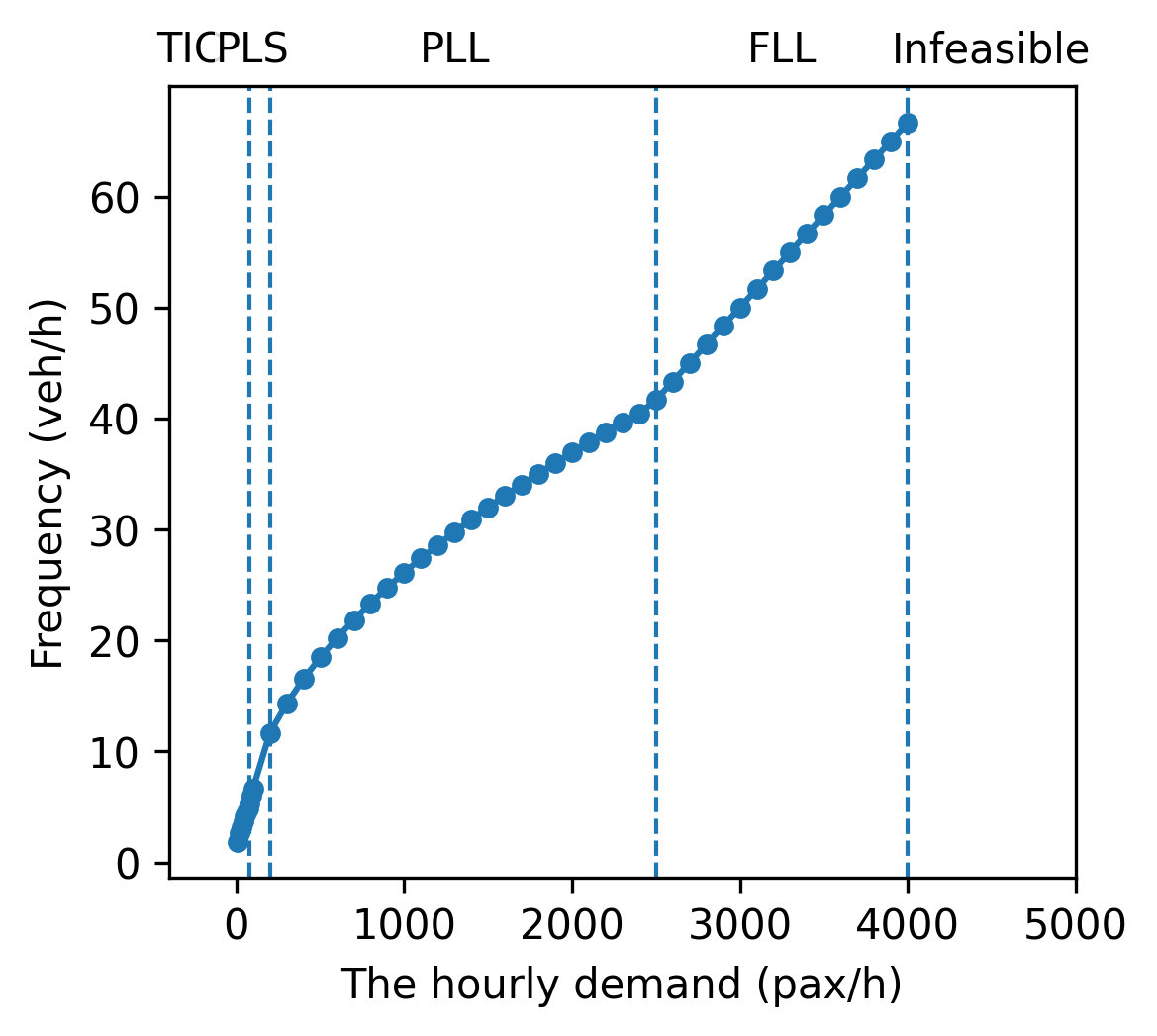}
        \caption{Relationship between optimal bus frequency and demand}
        \label{fig:small_opt_frequency}
    \end{subfigure}
    \hfill
    \begin{subfigure}[t]{0.45\linewidth}
        \centering
        \begin{picture}(0,0)
        \end{picture}
        \includegraphics[width=\linewidth]{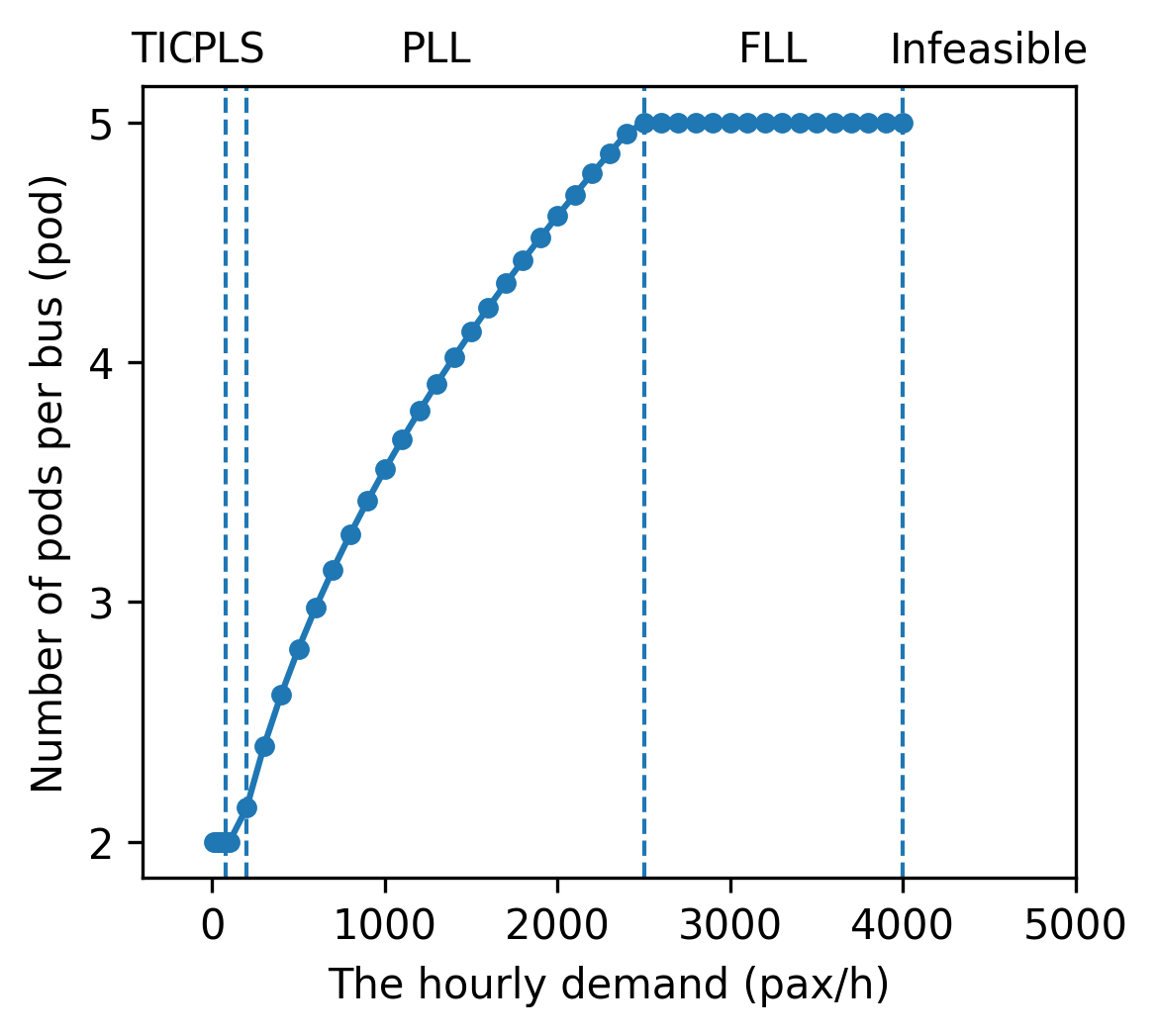}
        \caption{Relationship between optimal bus capacity (in terms of number of pods) and demand}
        \label{fig:small_opt_cap}
    \end{subfigure}
    \hfill

    \caption{The behavior of the optimal design for SLAM bus service with small pods (6 seats per pod) (a) frequency, (b) bus capacity}
    \label{fig:opt_small}
\end{figure}

\subsubsection{Case of large pods with capacity of 16 seats}
As pod capacity increases, the boarding and alighting time required by each boarding pod increases, which raises the minimum feasible headway and makes the MFH regime more likely to emerge. To capture this influence, the pod capacity is set to 16 seats in this case. As shown in Figure~\ref{fig:opt_core}, when the demand increases, the optimal frequency and number of pods follow the patterns below:

Under low ridership, the ``Through Idle Capacity'' regime emerges. In this regime, the optimal frequency increases in proportion to the square root of the demand, while the modular bus maintains a minimum of two pods (one non-boarding pod and one boarding pod) to enable non-stopping operations.

As the demand continues to grow, the ``Partially Loaded Small Buses'' regime arises. Here, the optimal frequency increases linearly with the demand, while the number of pods per bus is still fixed at the minimum of two. Compared to the ``Through Idle Capacity'' regime, the slope of the frequency-demand relationship becomes steeper, indicating that each additional passenger requires a larger increase in service frequency.

With the further increase in the ridership, the ``Partially Loaded Large Buses'' regime emerges, as maintaining the bus capacity at the minimum is no longer optimal. In this regime, both frequency and bus capacity increase with demand. Frequency grows more steeply than in the ``Through Idle Capacity" regime, while the bus capacity increases with the square-root of the demand.

Once the optimal frequency reaches the upper bound imposed by the minimum headway constraint, capacity expansion becomes the sole adjustment approach. In the MFH regime, bus capacity grows linearly with demand while frequency remains fixed at its maximum.

We find four regimes of the optimal design with large pods. Note that the Fully Loaded Large Buses regime does not appear in the experiment because the MFH threshold is reached before the FLL threshold. Initially, buses have the minimum of two pods and the frequency follows first the square-root and then linear growth with the demand. Subsequently, the frequency follows a steeper square-root growth and then remains fixed at its maximum, while bus capacity increases first with the square-root of the demand and later linearly with the demand. 

\begin{figure}[H]
    \centering
    \begin{subfigure}[t]{0.45\linewidth}
        \centering
        \begin{picture}(0,0)
        \end{picture}
        \includegraphics[width=\linewidth]{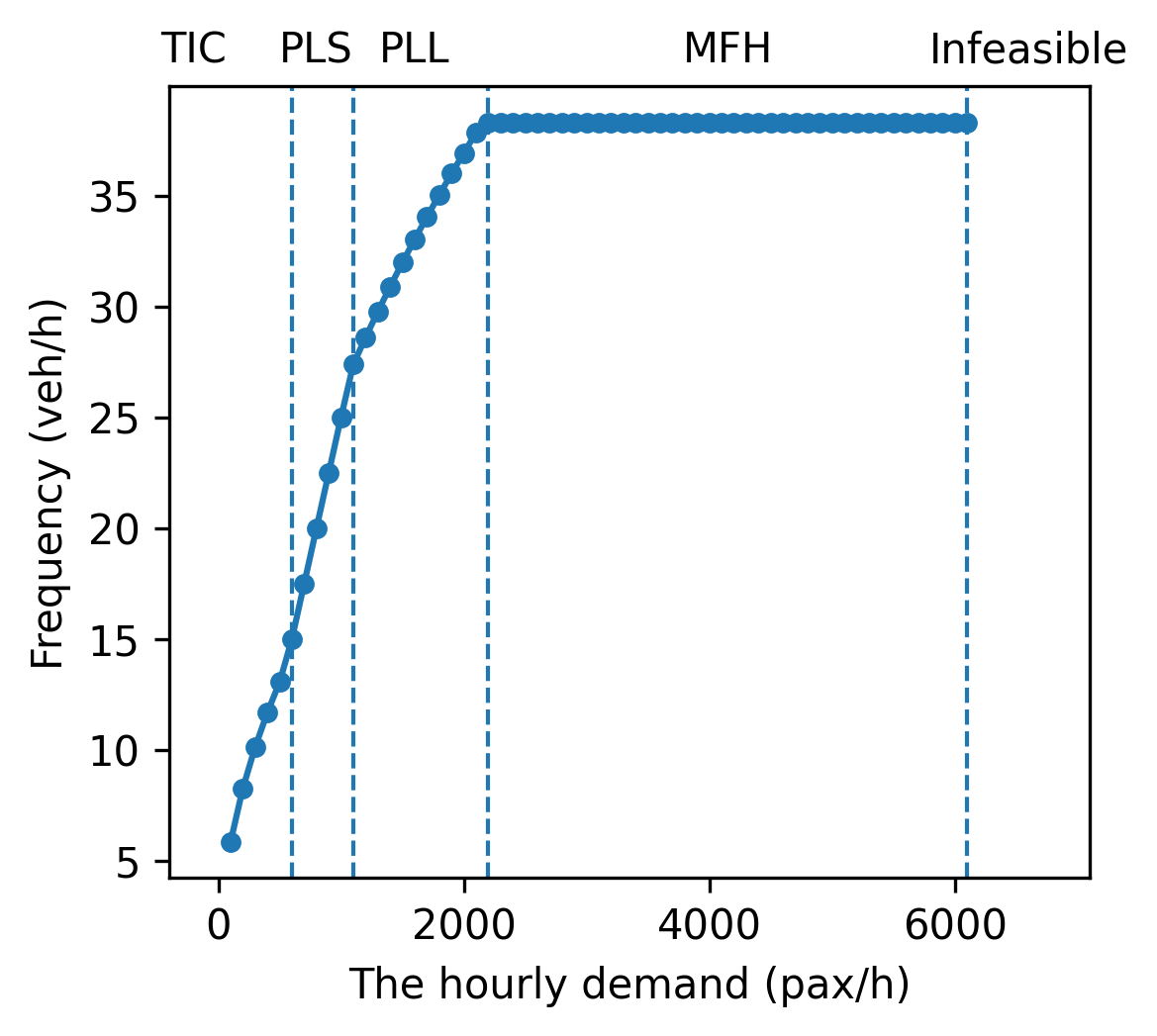}
        \caption{Relationship between optimal bus frequency and demand}
        \label{fig:core_opt_frequency}
    \end{subfigure}
    \hfill
    \begin{subfigure}[t]{0.45\linewidth}
        \centering
        \begin{picture}(0,0)
        \end{picture}
        \includegraphics[width=\linewidth]{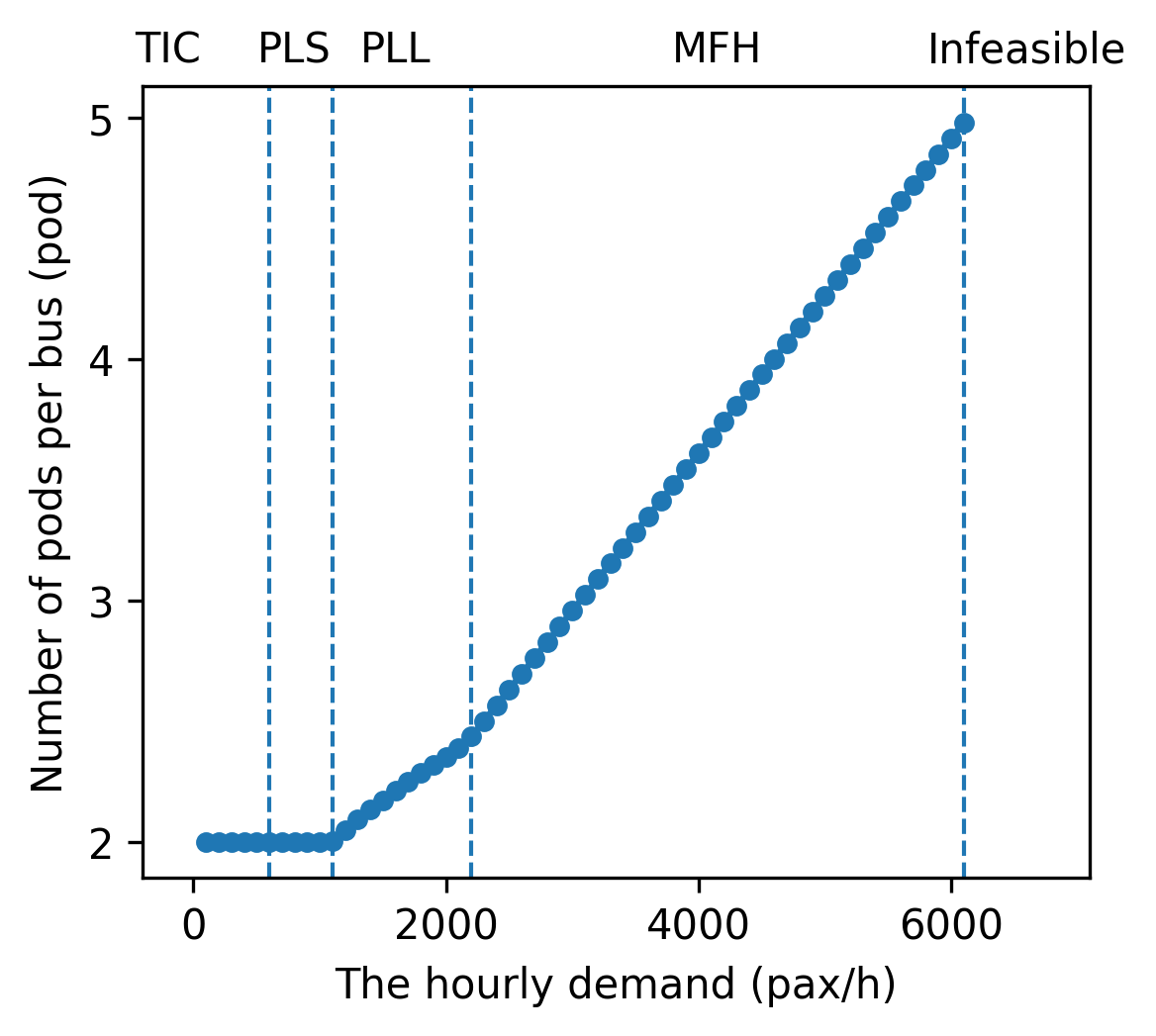}
        \caption{Relationship between optimal bus capacity (in terms of number of pods) and demand}
        \label{fig:core_opt_cap}
    \end{subfigure}
    \hfill

    \caption{The behavior of the optimal design for SLAM bus service with large pods (16 seats per pod) (a) frequency, (b) bus capacity}
    \label{fig:opt_core}
\end{figure}

\subsubsection{Numerical comparison with conventional buses}
We compare the total costs of SLAM and conventional buses when the optimal number of pods per bus is restricted to integer values. Based on the structure derived in the theoretical analysis, the integer pod-configuration solution is obtained by first solving the continuous model, then evaluating the two adjacent integer values of the number of pods per bus, and finally re-optimizing the frequency for each candidate. The design with the lower total cost is selected as the final solution. 
Following this, we compare the optimal design and total costs of SLAM and conventional buses. The results are illustrated in Figure~\ref{fig:AvgCostsComparison}.

In the case of small pods with 6 seats per pod, as shown in Figure~\ref{fig:small_opt_avg_costs}, we observe a sharp decline in average costs at low demand levels, followed by a more gradual decline in the PLL regime. This pattern is consistent with the theoretical analysis. That is, the average waiting costs decrease more slowly in the PLL regime compared to the preceding regimes. As a result, scale economies are weaker in the PLL regime. In the large pod case, as shown in Figure~\ref{fig:large_opt_avg_costs}, average costs always decrease, indicating the presence of scale economies, which are significantly stronger at low demand levels.

SLAM has higher average costs than conventional buses at low demand, especially with 6-seat pods. As demand increases, SLAM costs decline rapidly because vehicle and pod utilization improves. In both cases, SLAM becomes cost-competitive and cheaper than conventional buses at relatively high-demand levels within the feasible demand range. The cost advantage of SLAM is particularly strong for 16-seat pods.

\begin{figure}[H]
    \centering
    \begin{subfigure}[t]{0.45\linewidth}
        \centering
        \begin{picture}(0,0)
        \end{picture}
        \includegraphics[width=\linewidth]{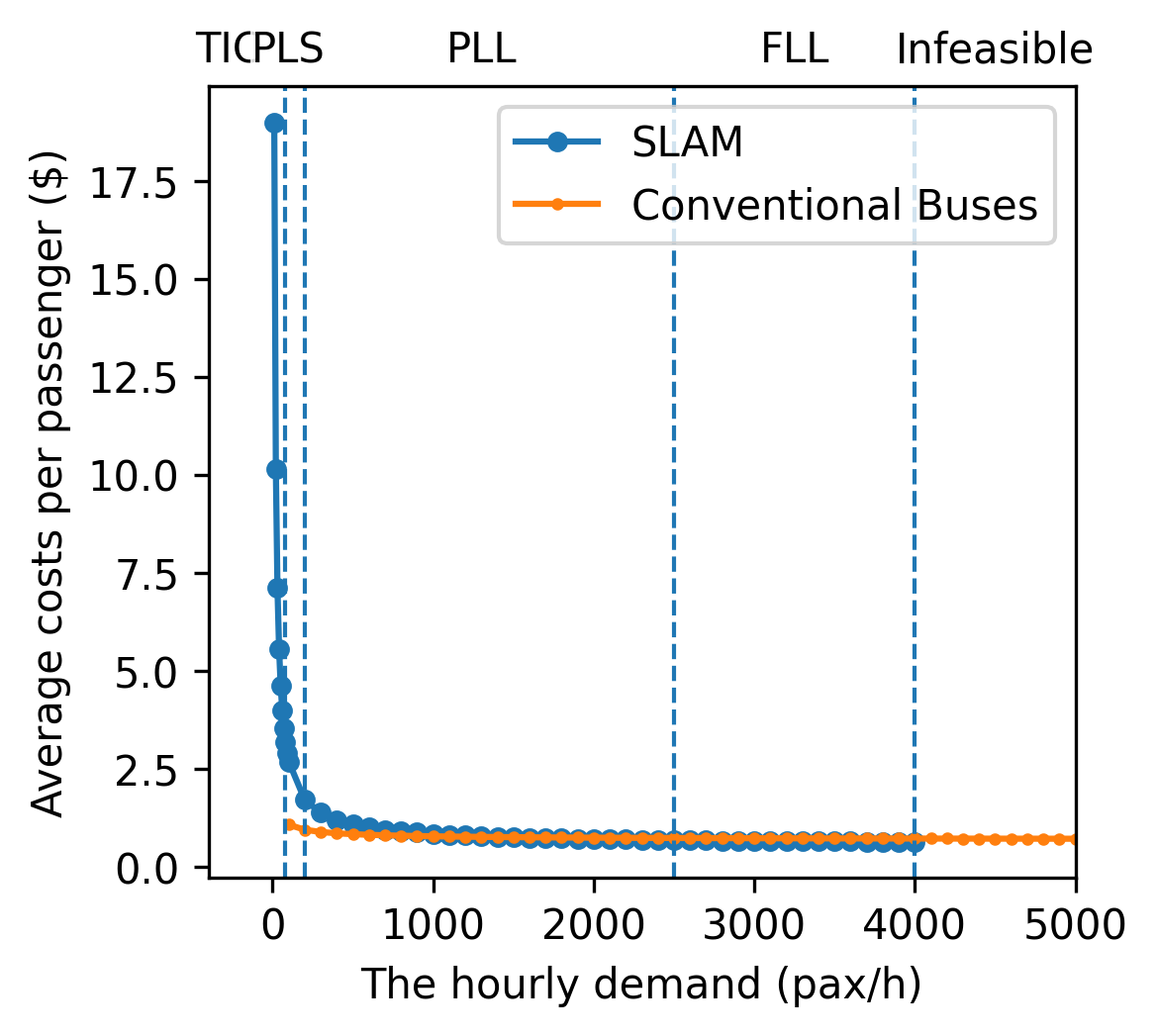}
        \caption{Relationship between demand and average cost per passenger for SLAM (6 seats per pod) and conventional buses}
        \label{fig:small_opt_avg_costs}
    \end{subfigure}
    \hfill
    \begin{subfigure}[t]{0.45\linewidth}
        \centering
        \begin{picture}(0,0)
        \end{picture}
        \includegraphics[width=\linewidth]{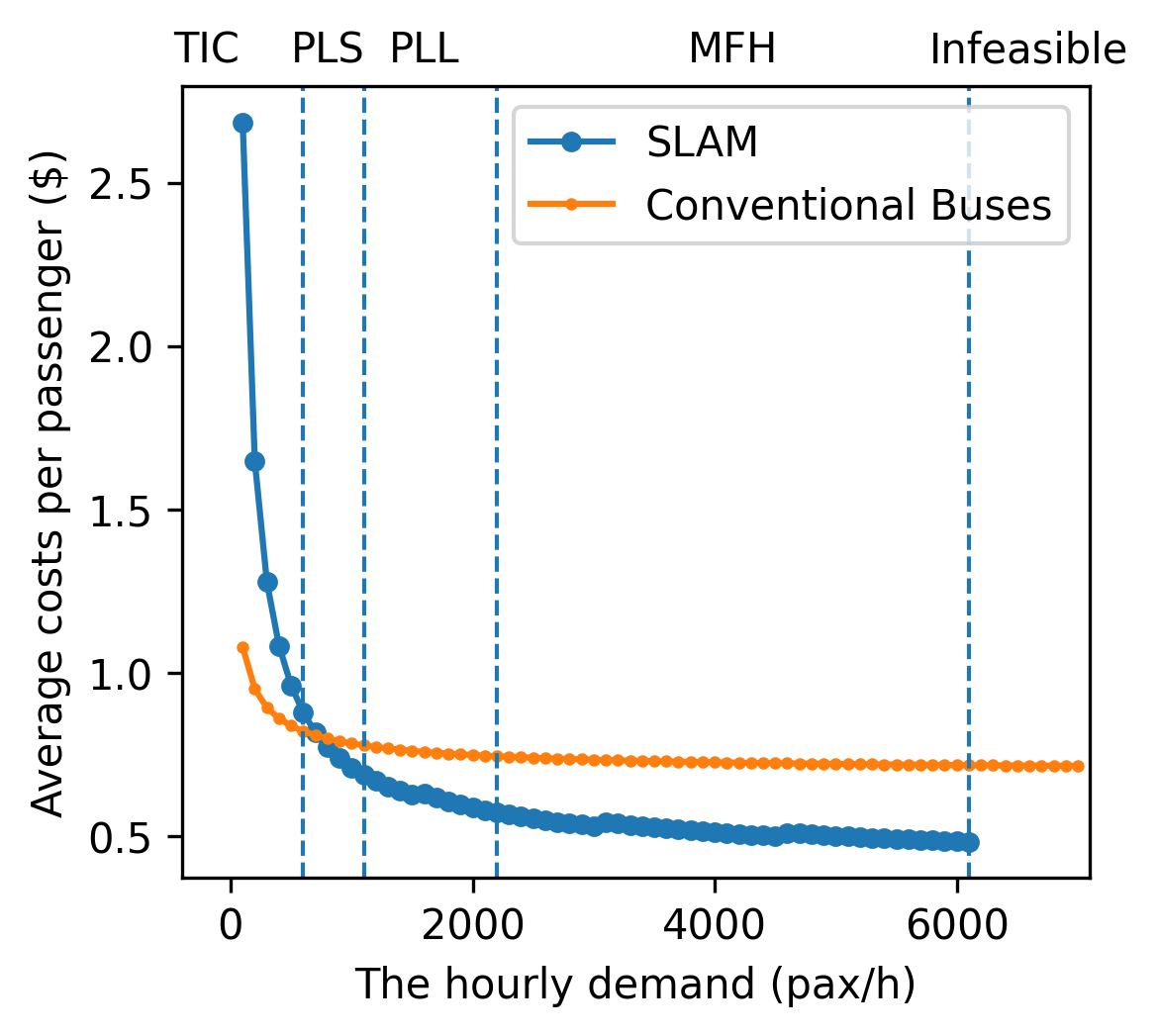}
        \caption{Relationship between demand and average cost per passenger for SLAM (16 seats per pod) and conventional buses}
        \label{fig:large_opt_avg_costs}
    \end{subfigure}
    
    \caption{Average costs per passenger under SLAM and conventional bus services. Panel (a) shows SLAM with 6-seat pods, and panel (b) shows SLAM with 16-seat pods. Dashed vertical lines indicate transitions between operating regimes and the demand level beyond which SLAM becomes infeasible.}
    \label{fig:AvgCostsComparison}
\end{figure}

\subsubsection{Discussion}
In the conventional model, higher demand leads to a higher optimized frequency and a larger vehicle capacity. Here, vehicle capacity refers to the planning-stage choice of vehicle type for a given demand level; once deployed, conventional buses operate with fixed capacity and cannot adjust capacity dynamically during service. In contrast, the SLAM bus service exhibits a different progression. First, only the optimal frequency increases with the ridership, while the optimal number of pods remains at its minimum. This is followed by an intermediate region in which both the frequency and the number of pods (i.e., bus capacity) increase simultaneously. Finally, when pod capacity is large, the number of pods continues to expand, while the frequency is capped at its maximum; otherwise, the number of pods remains fixed while the frequency continues to rise. 

Compared with the small-pod case, the SLAM bus service with large pods has a longer minimum feasible headway, which limits the optimal frequency to its feasible maximum. Note that the only difference between large-pod and small-pod experiments is the pod capacity, with all other parameters held constant. Therefore, in the large-pod case, the Minimum Feasible Headway regime becomes dominant and the longest regime, in contrast to the small-pod case, where the Partially Loaded Large Buses regime is the most prominent.

With a larger pod capacity, the feasible region for the non-stopping operation in the SLAM bus service is expanded. However, a larger pod capacity also reduces the maximum frequency, which in turn caps the optimal frequency at its maximum when the ridership is high.

The comparison with conventional buses shows that SLAM is less cost-effective at low demand because standby pods and the minimum two-pod configuration create additional operating costs. As demand increases, these fixed and capacity-related costs are spread over more passengers, and SLAM becomes more competitive. This effect is stronger with larger pods, for which SLAM can become cheaper than conventional buses over part of the feasible demand range. Overall, the results indicate that SLAM is most attractive at intermediate to relatively high demand levels, rather than at very low demand or beyond the feasibility limit.

\subsection{Sensitivity Analysis}
We conduct sensitivity analyses to examine how the optimal design and costs respond to changes in two SLAM-specific factors, i.e., coupling/decoupling time ($\theta$) and pod capacity ($K_P$). Figure~\ref{fig:sensitivity_KP_theta} reports the resulting optimal frequency, number of pods per bus, and average cost per passenger when $\theta$ varies from 30 s to 120 s and $K_P$ varies from 4 to 20 seats per pod, with hourly demand fixed at 3000 passengers. The default values of $\theta$ and $K_P$ are 30 s and 16 seats per pod, respectively. 

As $K_P$ increases, the average cost per passenger decreases because larger pods provide more capacity and reduce the required number of pods per bus. With fixed demand, as the pod capacity increases, the optimal design transitions from the FLL regime to the PLL regime and then to the MFH regime. The optimal number of pods declines with $K_P$, while the optimal frequency generally decreases for larger pod capacities. 

Increasing $\theta$ raises the minimum feasible headway and therefore restricts the maximum feasible frequency. Under the parameter values used in the experiment, the optimal design remains in the MFH regime. As a result, the optimal continuous frequency decreases with $\theta$, while the required number of pods increases. The average cost per passenger also increases slightly as coupling/decoupling becomes slower. For the integer solution, the optimal design changes in discrete steps, but it follows the same overall trends in frequency, pod number, and average cost.

The integer solution closely follows the continuous solution, although the discrete pod constraint leads to visible jumps in frequency and pod number. As discussed in section 3, the optimal integer value of $P$ is always either the floor or the ceiling of the continuous one. The average costs of the system are remarkably similar, showing that the closed-form solutions provide reliable guidance for SLAM planning.

Overall, the sensitivity analysis supports the robustness of the analytical framework, particularly in capturing the behavior of the optimal design. 


\begin{figure}[H]
    \centering

    \begin{subfigure}[t]{0.31\linewidth}
        \centering
        \includegraphics[width=\linewidth]{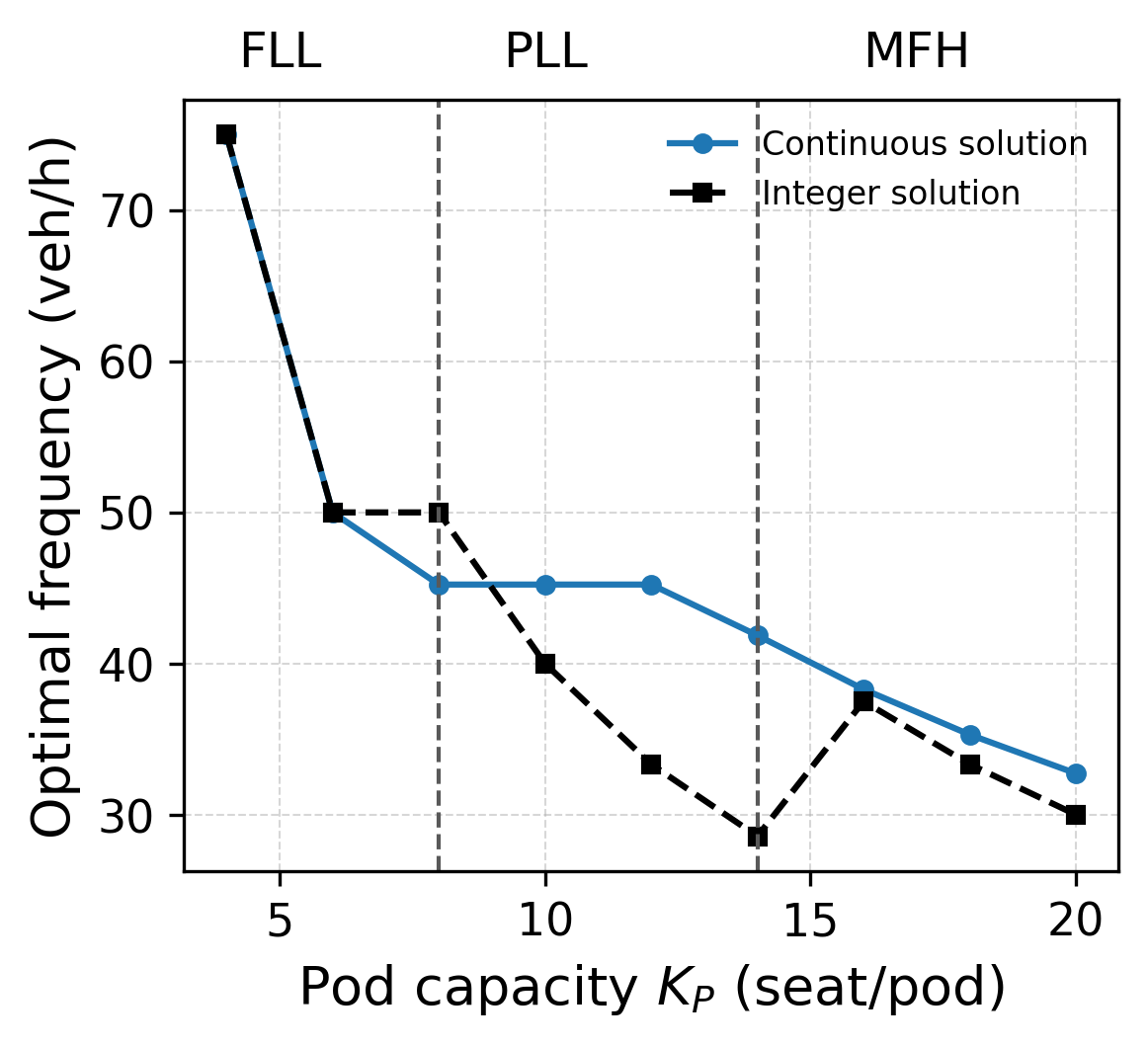}
        \caption{Frequency under varying $K_P$}
        \label{fig:sens_KP_frequency}
    \end{subfigure}
    \hfill
    \begin{subfigure}[t]{0.31\linewidth}
        \centering
        \includegraphics[width=\linewidth]{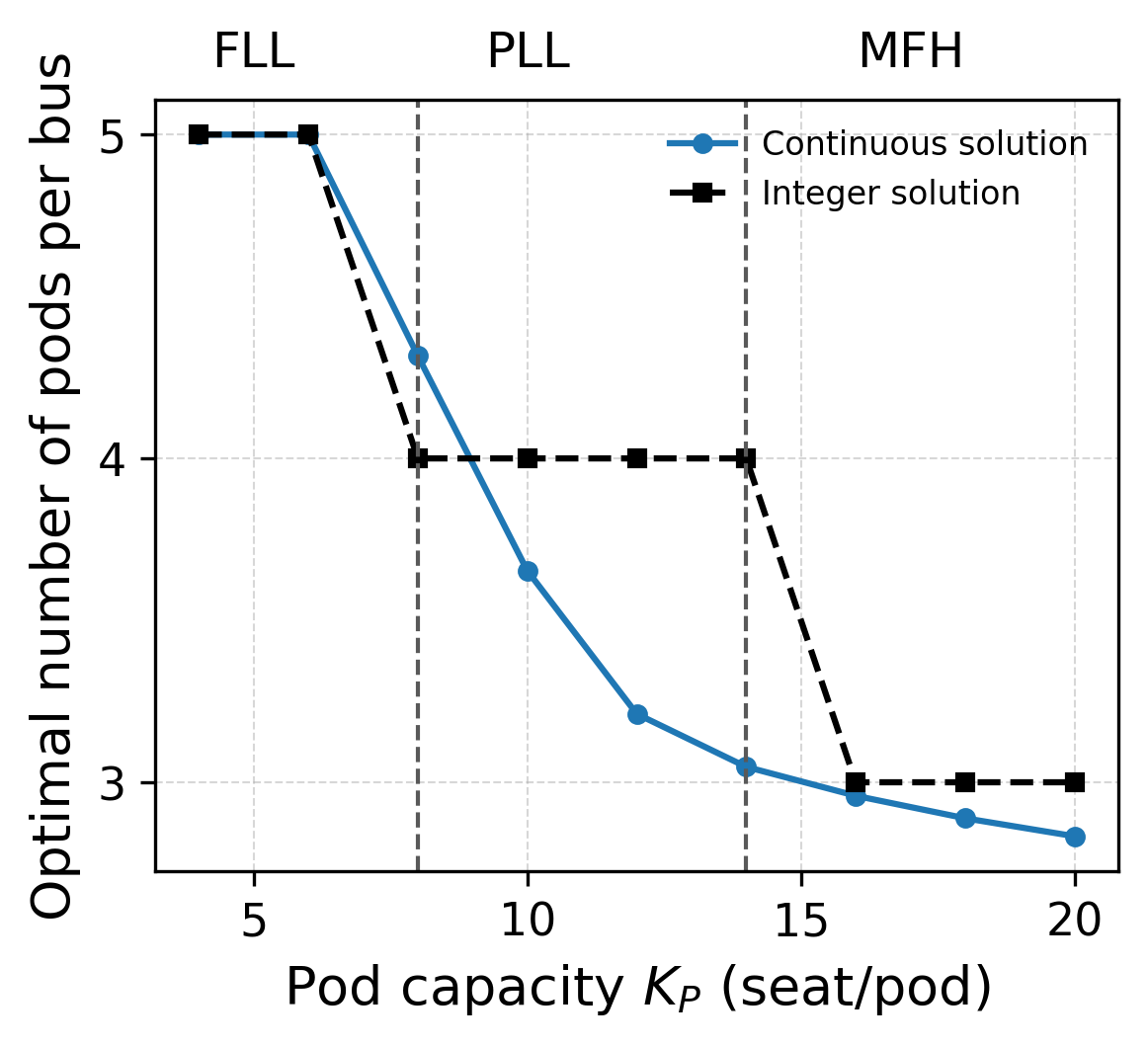}
        \caption{Number of pods under varying $K_P$}
        \label{fig:sens_KP_pods}
    \end{subfigure}
    \hfill
    \begin{subfigure}[t]{0.31\linewidth}
        \centering
        \includegraphics[width=\linewidth]{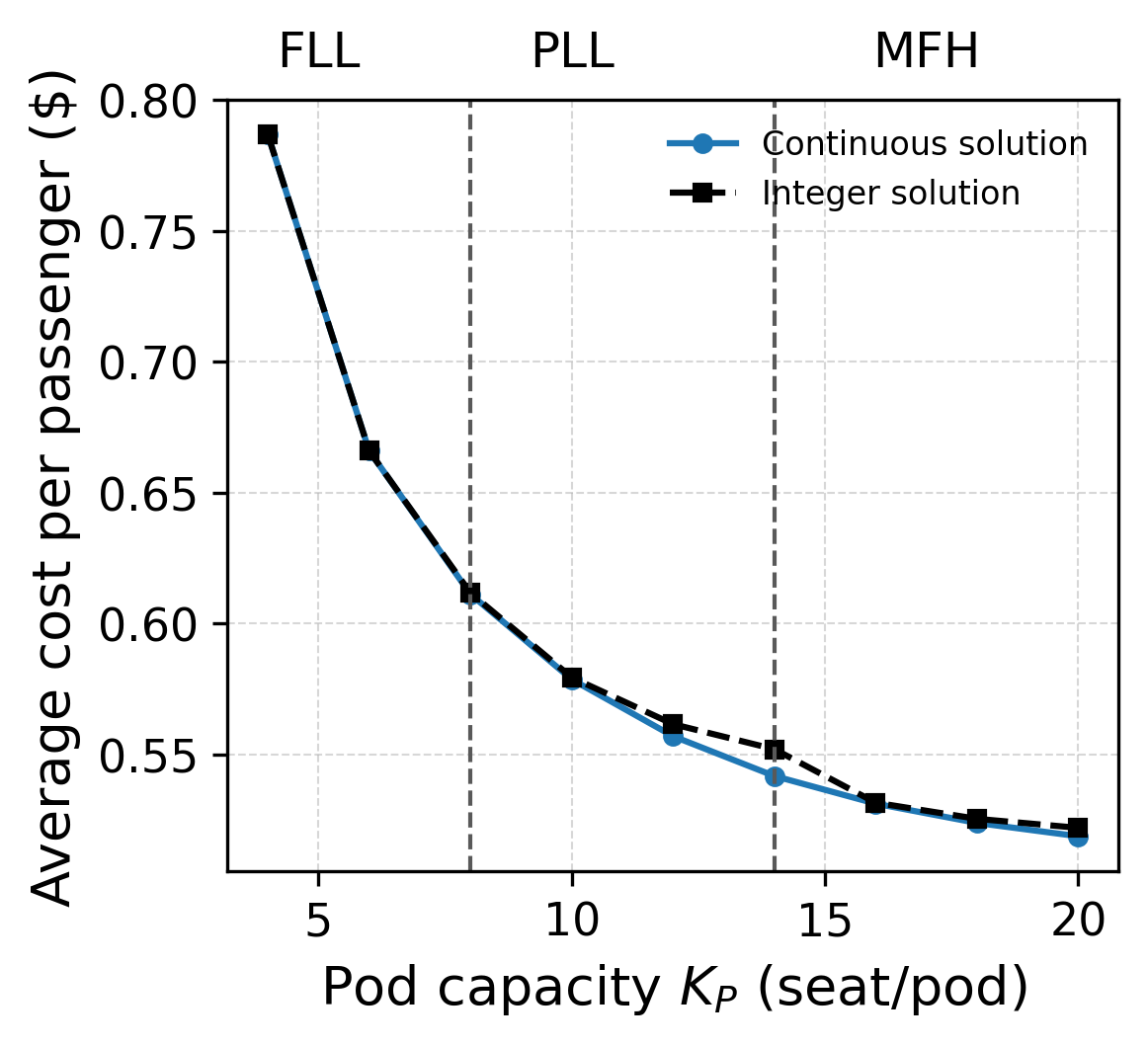}
        \caption{Average cost under varying $K_P$}
        \label{fig:sens_KP_cost}
    \end{subfigure}

    \vspace{0.5em}

    \begin{subfigure}[t]{0.31\linewidth}
        \centering
        \includegraphics[width=\linewidth]{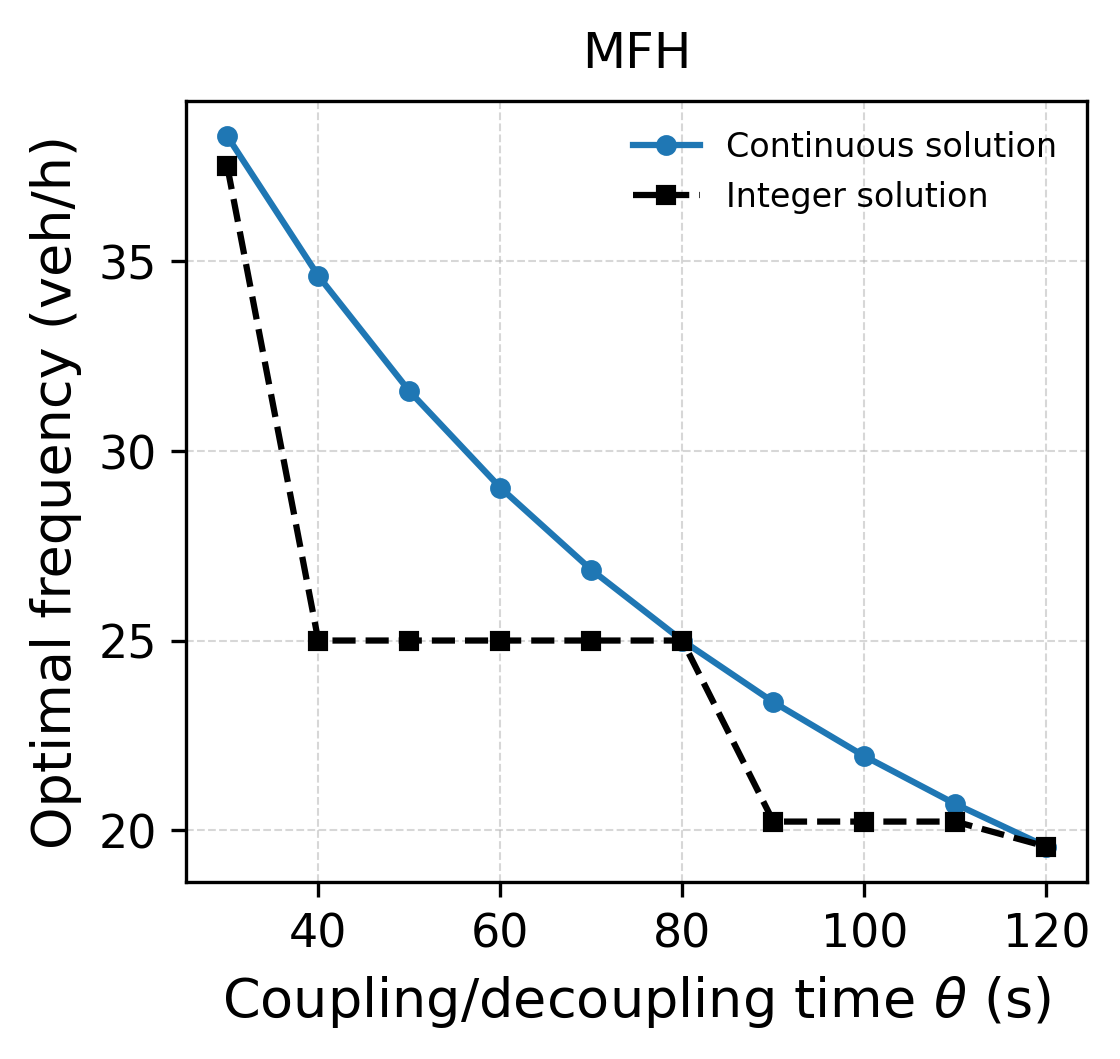}
        \caption{Frequency under varying $\theta$}
        \label{fig:sens_theta_frequency}
    \end{subfigure}
    \hfill
    \begin{subfigure}[t]{0.31\linewidth}
        \centering
        \includegraphics[width=\linewidth]{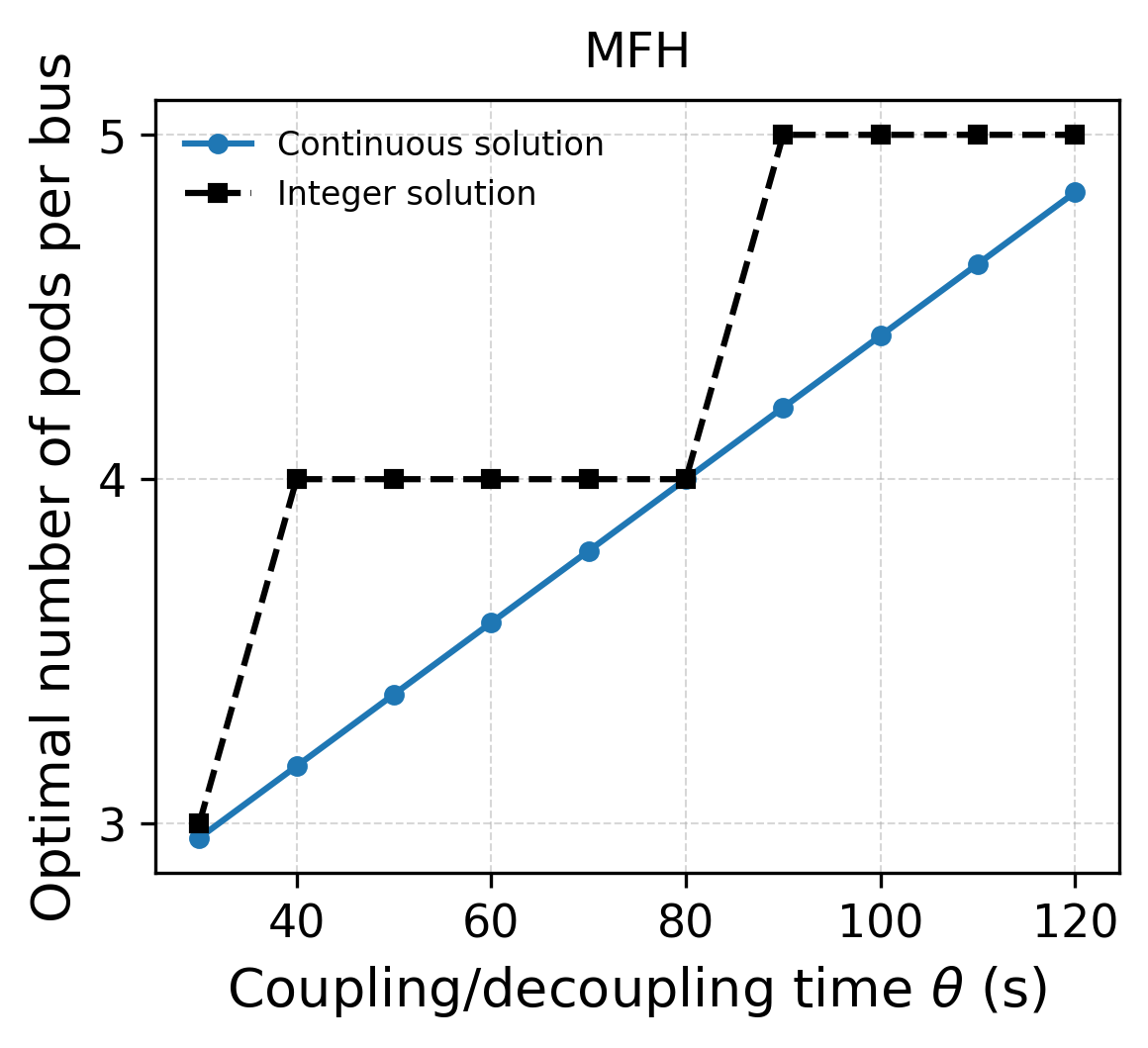}
        \caption{Number of pods under varying $\theta$}
        \label{fig:sens_theta_pods}
    \end{subfigure}
    \hfill
    \begin{subfigure}[t]{0.31\linewidth}
        \centering
        \includegraphics[width=\linewidth]{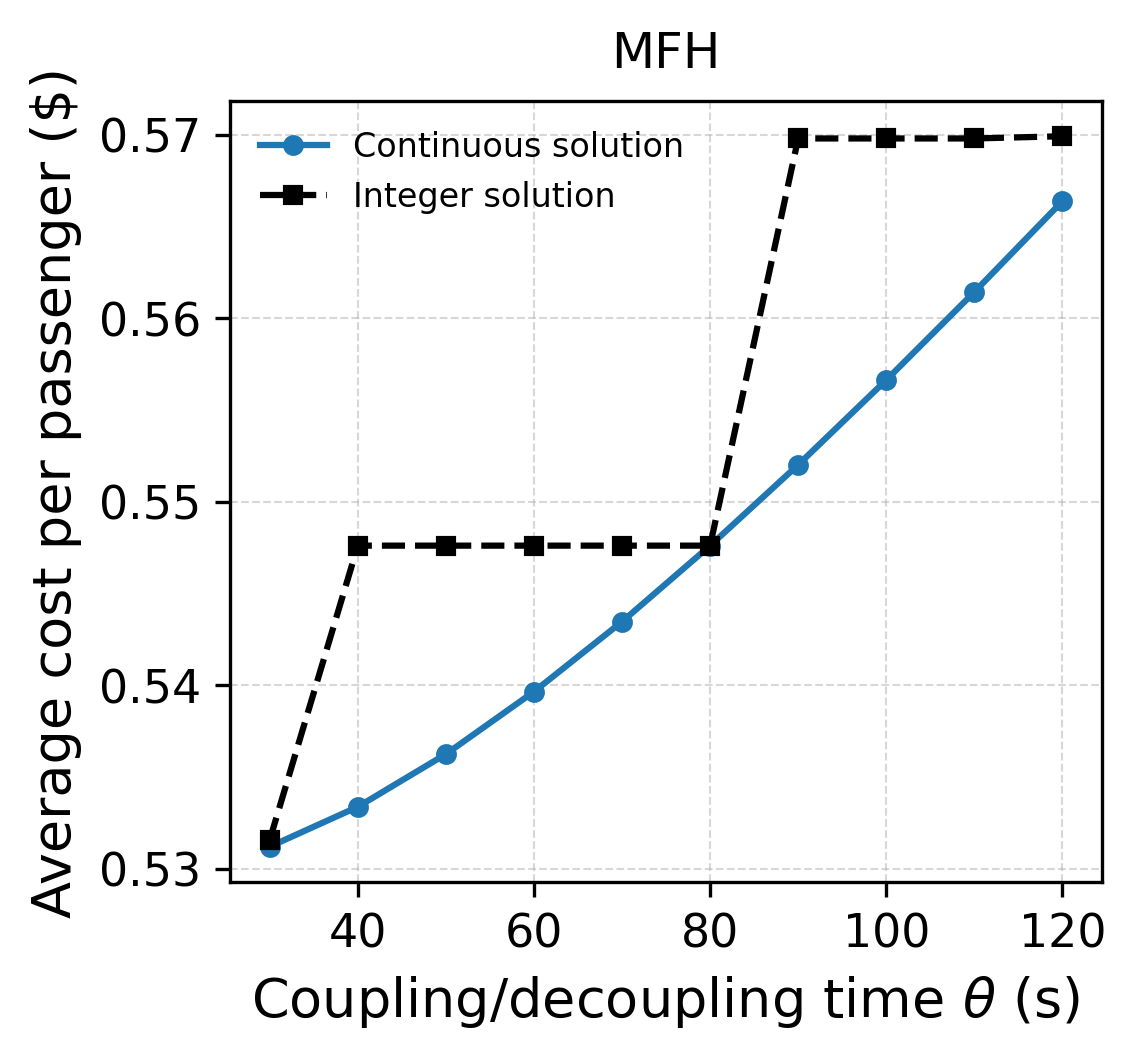}
        \caption{Average cost under varying $\theta$}
        \label{fig:sens_theta_cost}
    \end{subfigure}

    \caption{Sensitivity analysis of the optimal SLAM design with respect to pod capacity ($K_P$) and coupling/decoupling time ($\theta$), with hourly demand fixed at 3000 passengers. Solid blue lines represent continuous solutions, while black dashed lines represent integer solutions.}
    \label{fig:sensitivity_KP_theta}
\end{figure}



\section{Conclusions}
Modular autonomous buses have attracted considerable attention from researchers because they offer greater operational flexibility than conventional buses. Among their practical applications, the SLAM bus service is noteworthy, which detaches a pod to serve boarding and alighting and allows the remaining pods to continue without stopping. This service has been shown to reduce passengers' in-vehicle travel time by eliminating the dwell of through passengers. In this paper, we have proposed an analytical framework to optimize the SLAM system and further analyze its optimal solutions, feasibility conditions, and sources of scale economies.

Our model demonstrates that SLAM operation requires three novel mathematical constraints: a maximum headway, a minimum headway, and a minimum bus length constraint. The maximum headway constraint ensures that all passengers arriving within a headway interval can be accommodated in the boarding pod when performing non-stopping operations. The minimum headway constraint ensures that the boarding pod has enough time to perform these operations before attaching to the next bus. The minimum bus length constraint guarantees the bus has sufficient pods (at least two) to perform non-stopping operations. These constraints are novel in the SLAM bus service compared to conventional bus service.

Both the theoretical and numerical analyses characterize a set of mutually exclusive and collectively exhaustive regimes governing the optimal design as demand increases. First, the ``Through Idle Capacity'' regime arises when the ridership is low and buses are never full. In this regime, the optimal frequency increases with the square root of the demand and buses have two pods. Second, with the increase in the ridership, ``Partially Loaded Small Buses" regime occurs as the maximum load determines the frequency and bus capacity, in which the optimal frequency increases linearly with demand, and buses remain at a minimum length of two pods. Third, ``Partially Loaded Large Buses" regime appears when the ridership continues to grow. In this regime, both frequency and number of pods per bus increase with the square root of ridership. Fourth, as the ridership continues to increase, the ``Fully Loaded Large Buses" regime takes place. In this regime, the optimal frequency increases linearly with the ridership, while the bus capacity is determined by the peak segment-stop load ratio; a higher ratio indicates longer average travel distances and stronger through-flows. Finally, the minimum headway constraint may trigger the MFH regime midway through the sequence, in which the optimal frequency reaches its upper bound and the number of pods per bus grows proportionally to the demand.

Moreover, we highlight that the non-stopping operations become infeasible at high ridership levels. The reason is that the service frequency required to accommodate the accumulated ridership at some bus stations exceeds the maximum frequency compatible with the non-stopping operation. At those stations, the bus should do a full stop instead of SLAM. Our method identifies such stops, making it possible to design a hybrid system. 

When compared to conventional buses, we show that the SLAM bus service leads to longer waiting times under low demand due to its minimum bus length constraint, and non-stopping operation becomes infeasible at high demand levels. Consequently, the SLAM bus service is particularly appropriate for intermediate demand levels.

We identify four sources of scale economies in the SLAM bus service: the Mohring effect, through-capacity economies, boarding-capacity economies and standby-pod costs. The Mohring effect refers to the phenomenon in which rising ridership leads to higher service frequency, thereby reducing (average) passengers’ waiting times; this arises in regimes where frequency increases with demand, linearly or sublinearly. Through-capacity economies increase seat occupancy as ridership grows, spreading sublinearly increasing operators' cost over linearly increasing passengers. Boarding-capacity economies arise when additional boarding passengers can be accommodated by existing boarding-pod capacity. In the PLL and MFH regimes, these costs grow sublinearly or remain constant relative to ridership. As a result, a growing number of passengers share a diminishing incremental cost of non-stopping operations. Finally, standby pods are additional pods deployed at every stop to enable non-stopping operations. Their acquisition and operating costs are essentially fixed and determined by the number of bus stations. As more riders use the service, these fixed standby-pod costs are distributed across a growing passenger base.

The SLAM bus service outperforms conventional buses in terms of reducing in-vehicle time for through passengers. Whereas most papers on the SLAM bus are based on simulations and consider a fixed fleet and capacity, we formulate a novel analytical model to design the SLAM bus service. The paper demonstrates contributions in three main aspects: choosing the optimal frequency and number of pods per bus across various ridership levels, assessing feasibility of the SLAM service, and identifying sources of scale economies. The resulting findings provide guidelines for choosing the frequency and number of pods per bus, help operators determine which stop the non-stopping operation is feasible, and enable operators to deliver an economically sustainable service.

The analytical tractability of the present model relies on focusing on a single bus line. However, this single-line framework does not capture in-motion transfers at the network level. Incorporating this feature would increase model complexity and might reduce the analytical tractability needed to clearly characterize the optimal design, feasibility conditions, and economies of scale. Nevertheless, extending this framework to network-level settings remains an important direction for future research. Another important direction is to incorporate stochastic and time-varying demand, which would require adaptive control strategies. In this case, service frequency and pod allocation would be adjusted in real time in response to observed demand and operating conditions. 

Our results suggest that SLAM is most promising at intermediate demand levels rather than under very high demand. This implies that SLAM may be less suitable for the busiest corridors. For a bus line with both high-demand stops and moderate-demand stops, a hybrid operation may be appropriate: SLAM can be used at moderate-demand stops, while high-demand stops can be served through conventional stopping operations. Such a design could preserve part of the travel-time benefits of SLAM while maintaining feasibility at heavily used stops.

The scale economies in SLAM encourage operators to increase frequency, thereby reducing passenger waiting time, attracting more passengers, and creating a virtuous cycle for sustainable operations \citep{gao_roadspace_2026, bar-yosef_model_2013}. At low demand levels, however, neither conventional buses nor SLAM may be optimal, indicating that alternative operational strategies for modular buses should be explored, such as using them to provide on-demand mobility \citep{ye_integrating_2025, luo_optimal_2025}. These findings provide guidance for policymakers on where SLAM deployment is most effective and highlight the need for further research on flexible service designs under low-demand conditions.


\appendix
\renewcommand{\thesection}{\appendixname~\Alph{section}}
\renewcommand{\thefigure}{\thesection.\arabic{figure}}
\setcounter{figure}{0}

\section{Comparison between the SLAM and conventional bus service in terms of the total costs and outperforming ridership ranges of the SLAM bus service}
\label{appdx:Comp_low_demand}

To compare the total costs of the conventional bus service and the SLAM bus service, we first derive the difference between their total cost functions and then identify the condition under which the SLAM bus service becomes more economical.

 We replace the frequency in Equation~\eqref{obj:trad_single} with Equation~\eqref{Eq:OptFreqTrad} and the minimized total costs of the conventional bus service are 
\begin{equation}\label{Eq:OptConvCosts}
C^*_{CONV}
=
2\sqrt{
(T + S \tau)\gamma_0
\left(
\frac{\pi_w X}{2}
+\pi_v\frac{l}{L}tX^2
+t\gamma_1\rho_{\max}X^2
\right)
}
+\pi_v\frac{l}{L}(T + S \tau)X
+t\gamma_0X
+(T + S \tau)\gamma_1\rho_{\max}X
\end{equation}

We subtract the optimized total costs of SLAM in the TIC regime, given in Equation~\eqref{Eq:CostsIC}, from the optimized costs of the conventional buses in Equation~\eqref{Eq:OptConvCosts}, which yields 
\begin{equation}\label{eq:delta_cost_grouped}
\begin{aligned}
&\Delta C_{TIC}(X)= C^*_{CONV} - C^*_{TIC} \\
&= 2
\left[
\sqrt{
(T+S\tau)\gamma_0
\left(
\frac{\pi_w X}{2}
+
tX^2\left(\pi_v\frac{l}{L}+\gamma_1\rho_{\max}\right)
\right)
}
-
\sqrt{\pi_w T\gamma X}
\right]
+
X\left[
\pi_v\frac{l}{L}S\tau+t\gamma_0+(T+S\tau)\gamma_1\rho_{\max}
\right]-S\gamma 
\end{aligned}
\end{equation}

When demand is sufficiently low, the cost gap between the two optimized designs is negative. Specifically, since $\Delta C_{TIC}(X = 0) = -S \gamma < 0$ and $\Delta C_{TIC}(X)$ is a continuous function on $X$, there exists a non-empty interval of low demand levels for which $\Delta C_{TIC}(X) < 0$. In this interval, conventional buses have lower costs than SLAM.

At low demand levels, the boarding and alighting process adds only a small amount to in-vehicle time, so the time-saving benefits of deploying standby pods are limited. In this case, the additional cost of maintaining standby pods outweighs the reduction in in-vehicle time. For a given fleet size, it is therefore more cost-effective to operate pods individually and increase service frequency, thereby reducing passenger waiting time, than to run two coupled pods with standby support under SLAM. Hence, SLAM is not cost-effective at low demand levels.

\newpage
\bibliographystyle{apalike}  
\bibliography{modular_bus, single_line}

\end{document}